\documentclass{aa}  

\usepackage{graphicx}
\usepackage{txfonts}
\usepackage{lipsum}
\usepackage{subcaption}         % necessary for continued figures
\usepackage{lscape}             % to rotate a single page table
\usepackage{placeins}           % useful with \FloatBarrier, to keep 
\usepackage{xcolor}
\usepackage{lipsum}

\usepackage{hyperref}
\hypersetup{
    colorlinks=true,
    linkcolor=blue,
    citecolor=blue,
    filecolor=magenta,      
    urlcolor=blue,
}

\renewcommand\makeLineNumber{} % turn off linenumbers for arXiv

\begin{document}

%%%%%%%%%%%%%%%%%%%%%%%%%%%%%%%%%%%%%%%%%%%%%%%%%%%%%%%%%%%%%%%%%%%%%%%%%%%%%%%%%%%%%%%%%%
%%%%%%%%%%%%%%%%%%%%%%%%%%%%%%%%%%%%%%%%%%%%%%%%%%%%%%%%%%%%%%%%%%%%%%%%%%%%%%%%%%%%%%%%%%

\title{Icy or rocky? Convective or stable?}

\subtitle{New interior models of Uranus and Neptune}
\author{Luca Morf\thanks{Corresponding author, luca.morf@uzh.ch}
    \and Ravit Helled
    }

\institute{Department of Astrophysics, University of Zürich, Winterthurerstrasse 190, 8057 Zürich, Switzerland
         }

\date{Received 19 August 2025, Accepted 27 September 2025}

%%%%%%%%%%%%%%%%%%%%%%%%%%%%%%%%%%%%%%%%%%%%%%%%%%%%%%%%%%%%%%%%%%%%%%%%%%%%%%%%%%%%%%%%%%
%%%%%%%%%%%%%%%%%%%%%%%%%%%%%%%%%%%%%%%%%%%%%%%%%%%%%%%%%%%%%%%%%%%%%%%%%%%%%%%%%%%%%%%%%%

\abstract{
We present a new framework for constructing agnostic and yet physical models for planetary interiors and apply it to Uranus and Neptune. 
Unlike previous research that either impose rigid assumptions or rely on simplified empirical profiles, our approach bridges both paradigms. 
Starting from randomly generated density profiles, we applied an iterative algorithm that converges towards models that simultaneously satisfy hydrostatic equilibrium, match the observed gravitational moments, and remain thermodynamically and compositionally consistent.
The inferred interior models for Uranus and Neptune span a wide range of possible interior structures, in particular encompassing both water-dominated and rock-dominated configurations (rock-to-water mass ratios between 0.04--3.92 for Uranus and 0.20--1.78 for Neptune).
All models contain convective regions with ionic water and have temperature–pressure profiles that remain above the demixing curves for hydrogen–helium–water mixtures.
This offers both a plausible explanation for the observed non-dipolar magnetic fields and indicates that no hydrogen–helium–water demixing occurs. 
We find a higher H-He mass fraction in the outermost convection zones for Uranus (0.62--0.73) compared to Neptune (0.25--0.49) and that Uranus' magnetic field is likely generated deeper in the interior compared to Neptune.
We infer upper limits of 0.69--0.74 (Uranus) versus 0.78--0.92 (Neptune) for the outer edges of the dynamo regions in units of normalised radii.
Overall, our findings challenge the conventional classification of Uranus and Neptune as 'ice giants' and underscore the need for improved observational data or formation constraints to break compositional degeneracy.
}

%%%%%%%%%%%%%%%%%%%%%%%%%%%%%%%%%%%%%%%%%%%%%%%%%%%%%%%%%%%%%%%%%%%%%%%%%%%%%%%%%%%%%%%%%%
%%%%%%%%%%%%%%%%%%%%%%%%%%%%%%%%%%%%%%%%%%%%%%%%%%%%%%%%%%%%%%%%%%%%%%%%%%%%%%%%%%%%%%%%%%


%%%%%%%%%%%%%%%%%%%%%%%%%%%%%%%%%%%%%%%%%%%%%%%%%%%%%%%%%%%%%%%%%%%%%%%%%%%%%%%%%%%%%%%%%%
%%%%%%%%%%%%%%%%%%%%%%%%%%%%%%%%%%%%%%%%%%%%%%%%%%%%%%%%%%%%%%%%%%%%%%%%%%%%%%%%%%%%%%%%%%

\maketitle

%%%%%%%%%%%%%%%%%%%%%%%%%%%%%%%%%%%%%%%%%%%%%%%%%%%%%%%%%%%%%%%%%%%%%%%%%%%%%%%%%%%%%%%%%%
%%%%%%%%%%%%%%%%%%%%%%%%%%%%%%%%%%%%%%%%%%%%%%%%%%%%%%%%%%%%%%%%%%%%%%%%%%%%%%%%%%%%%%%%%%

\section{Introduction}
\label{sec:introduction}

%%%%%%%%%%%%%%%%%%%%%%%%%%%%%%%%%%%%%%%%%%%%%%%%%%%%%%%%%%%%%%%%%%%%%%%%%%%%%%%%%%%%%%%%%%

The outer solar system harbours intriguing, and yet the least understood planets in the Solar System. 
Uranus and Neptune, often referred to as 'ice giants', remain enigmatic in terms of their composition and formation and evolution history \citep[for example][and references therein]{Helled2025}. 
These two planets have always been key to understanding the solar system, but their importance has become even more apparent in recent years. 
Intermediate-mass exoplanets, with masses between Earth and Neptune, are found to be the most common planets in the galaxy \citep[for example][]{Zhu2021}. 
A better understanding of these giants could provide answers to several fundamental open problems in planetary physics. 
For example, it could reveal the mechanisms most relevant for the formation of a planet at a given distance from a specific star. 
It might help quantify the relationship between the observable atmospheres and the hidden interiors of a planet. 
And it could shed light on how a planet’s evolution after formation influences what we are able to observe today. 
To find answers to such problems, advanced planet formation and evolution models are needed.
Such models can be improved by obtaining constraints from planetary interior models.
In particular, it is important to ensure that such interior models are built with a minimal set of assumptions.
They should provide an unbiased view regarding the possible internal structures of the planets. 
In this work, we present a novel and agnostic method to generate self-consistent planetary interior models by improving over previous work \citep[][]{Morf2024}.

%%%%%%%%%%%%%%%%%%%%%%%%%%%%%%%%%%%%%%%%%%%%%%%%%%%%%%%%%%%%%%%%%%%%%%%%%%%%%%%%%%%%%%%%%%

Interior models are designed to fit the measured physical properties of a planet. 
These correspond to the planetary mass $M$, equatorial radius $R_\text{eq}$, rotation period $P_\text{rot}$, and external (Newtonian) gravitational field $V_\text{ext}$: 
\begin{equation}
V_\text{ext}(\vec{r}) = \frac{GM}{r}\left( 1 - \sum_{n=1}^{\infty} \left( \frac{R_\text{ref}}{r} \right)^{2n} J_{2n} P_{2n}(\cos\vartheta) \right), 
\end{equation}
expressed by gravitational moments $J_{2n}$. 
We assume a north-south and azimuthally symmetric planet described by spherical coordinates $\vec{r} = (r,\vartheta,\varphi)$. 
The Legendre polynomials are denoted as $P_{2n}$ and appear during the expansion of $V$ using spherical harmonics \citep[for example][]{Zharkov1978}. 
The reference radius $R_\text{ref}$ is an arbitrary normalisation constant that can be equal to $R_\text{eq}$, but does not have to be. 
The gravitational moments $J_{2n}$ have only been measured for planets within our solar system, where Voyager 2 \citep[for example][]{French1988} provided the first data for Uranus and Neptune. 
These values have later been improved using moon and ring observations \citep[][]{Jacobson2009, Jacobson2014, Wang2023, French2024, Jacobson2025}. 
In a corotating reference frame, the total exterior potential $U_\text{ext}$ is then given by:  
\begin{equation}
    U_\text{ext}(\vec{r}) = V_\text{ext}(\vec{r}) + Q(\vec{r}) = V_\text{ext}(\vec{r}) + \frac{1}{2}\left(\frac{2\pi}{P_\text{rot}}\right)^2r^2\sin^2\vartheta,
\end{equation}
assuming uniform rotation.

%%%%%%%%%%%%%%%%%%%%%%%%%%%%%%%%%%%%%%%%%%%%%%%%%%%%%%%%%%%%%%%%%%%%%%%%%%%%%%%%%%%%%%%%%%

\begin{table*}
\caption{
Physical data for the planets Uranus and Neptune that was used as a part of this work.
}
\centering
\begin{tabular}{lllll}
\hline
\hline
 & Uranus & & Neptune & \\
\hline
Mass $M$ [10$^{24}$ kg] & 86.8099 & \cite{Jacobson2025} & 102.409 & \cite{Jacobson2009} \\
Equatorial radius $R_\text{eq}$ [km] & 25559 & \cite{Lindal1987} & 24766 & \cite{Lindal1992} \\
Pressure $P$ at $R_\text{eq}$ [bar] & 1 & \cite{Lindal1987} & 1 & \cite{Lindal1992} \\
Reference radius $R_\text{ref}$ [km] & 25559 & \cite{French2024} & 25225 & \cite{Wang2023} \\
Rotation period $P_\text{rot}$ [s] & 62064 & \cite{Desch1986} & 57479 & \cite{Karkoschka2011} \\
Gravitational moment $J_2$ [$\cdot 10^6$] & 3509.291 $\pm$ 0.412 & \cite{French2024} & 3401.655 $\pm$ 3.994 & \cite{Wang2023}\\
Gravitational moment $J_4$ [$\cdot 10^6$] & -35.522 $\pm$ 0.466 & \cite{French2024} & -33.294 $\pm$ 10.000 & \cite{Wang2023} \\
\hline
\hline
\end{tabular}
\label{tab:measured_data}
\end{table*}

%%%%%%%%%%%%%%%%%%%%%%%%%%%%%%%%%%%%%%%%%%%%%%%%%%%%%%%%%%%%%%%%%%%%%%%%%%%%%%%%%%%%%%%%%%

We briefly summarise recent progress in interior modelling of Uranus or Neptune.
This includes work with so-called physical and empirical models.
Physical models are self-consistent and provide a complete picture of a planet's interior.
They include a density, pressure, temperature, and composition profile compatible with a given Equation of State (EoS).
However, physical models strongly depend on the assumptions used by the modeller.
They only provide a single and rather biased possibility among many into a given planet's interior.
Empirical models a priori provide only the pressure and density profile within a planet.
The interior temperature and composition can be inferred in a second step with an algorithm.
Since density and pressure were first inferred without any EoS knowledge, such algorithms typically cannot adhere to the same standards as used for physical modelling.
Empirical results for the temperature and composition hence suffer from inconsistencies or unphysical features \citep[for example][]{Neuenschwander2024, Morf2024}.
However, empirical models employ a minimal set of assumptions, enabling the modellers to achieve a more general view of the possible solutions for the planetary interior. 

%%%%%%%%%%%%%%%%%%%%%%%%%%%%%%%%%%%%%%%%%%%%%%%%%%%%%%%%%%%%%%%%%%%%%%%%%%%%%%%%%%%%%%%%%%

Recent work on physical interior models of Uranus or Neptune includes \cite{Militzer2024, CanoAmoros2024, Arevalo2025}.
\cite{Militzer2024} base their models on ab-initio simulations for a mixture of water, methane, and ammonia.
They show that these components phase separate under the pressure–temperature conditions in the interiors of Uranus and Neptune. 
The inferred interiors that match the observed gravity field and are compatible with magnetic field measurements are rock-poor:
2 \% and 7 \% of the total mass for Uranus and Neptune, respectively. 
\cite{CanoAmoros2024} provide hydrogen-water phase diagrams from available experimental and computational data. 
They find interior adiabatic structure models for Uranus and Neptune and infer a strong water depletion in the top layers of both planets.
This is caused by the immiscibility of hydrogen and water at certain conditions. 
\cite{Arevalo2025} present non-adiabatic and inhomogeneous evolution models for Uranus and Neptune assuming an interior composition of hydrogen, helium, methane, ammonia, water, and rocks. 
By setting the relevant initial parameters, they produce a Uranus model that preserves much of its primordial internal heat. 
In contrast, their Neptune model undergoes adiabatic cooling of the outer envelope, in agreement with observations.

%%%%%%%%%%%%%%%%%%%%%%%%%%%%%%%%%%%%%%%%%%%%%%%%%%%%%%%%%%%%%%%%%%%%%%%%%%%%%%%%%%%%%%%%%%

Recent work on empirical interior models of Uranus or Neptune includes \cite{Movshovitz2022, Neuenschwander2024, Malamud2024, Morf2024, Lin2025, Mankovich2025}.
\cite{Movshovitz2022} investigate the constraining power of a high-precision orbiter measuring the gravitational fields of Uranus and Neptune.
They find that while the ambiguity of Uranus' and Neptune's compositional profiles will still persist with high-precision data, determining the locations of any large composition gradient regions could become feasible.
\cite{Neuenschwander2024} model the density of Uranus with up to three polytropes.
Their models for Uranus’ interior include non-adiabatic regions.
This leads to significantly hotter internal temperatures and higher rock-to-water ratios compared to physical models. 
\cite{Malamud2024} show that Uranus and Neptune could have accreted refractory-dominated planetesimals, while still remaining icy. 
They investigate chemical reactions of organic-rich refractory materials and the hydrogen in gaseous atmospheres of protoplanets.
Uranus and Neptune could hence consist of large amounts of methane 'ice'. 
\cite{Morf2024} provide a random algorithm to interpret empirical models of Uranus in terms of their temperature and composition. 
Additionally, they supply the Theory of Figures \citep{Zharkov1978} to tenth order, allowing for accurate calculations of higher-order gravitational moments. 
Their models in particular constrain the properties of the ionic water convection zone, likely responsible for Uranus’ magnetic field.
\cite{Lin2025} argue that high degrees of mixing are required for Uranus interior models to be consistent with the measured gravitational moments. 
Furthermore, they show that the gravity measurements of a future orbiter could distinguish between high and low atmospheric metallicity scenarios. 
\cite{Mankovich2025} provide the spectra that could manifest in resonances with ring orbits or in Doppler imaging of Uranus’s visible surface. 
Their approach, originally developed for stellar oscillations, has the potential to significantly reduce the degeneracy in the set of otherwise allowed interior profiles.

%%%%%%%%%%%%%%%%%%%%%%%%%%%%%%%%%%%%%%%%%%%%%%%%%%%%%%%%%%%%%%%%%%%%%%%%%%%%%%%%%%%%%%%%%%

The aim of this study is to develop methods that generate interior models that are both agnostic and physical, to bridge the gap between physical and empirical modelling.
Our paper is organised as follows:
In Section \ref{sec:methods} we introduce our methods in more detail and give a general overview of our algorithm.
In Sections \ref{sec:Uranus} and \ref{sec:Neptune} we show the inferred interior models of Uranus and Neptune, respectively.
In Section \ref{sec:implications}, we elaborate and compare the inferred properties for Uranus and Neptune such as their hydrogen-helium abundances and rock-to-water ratios (Section \ref{sec:bulk_composition}).
We investigate the models' consistency with the observed magnetic fields (Section \ref{sec:magnetic_fields}) and immiscibility constraints from hydrogen-helium-water mixtures (Section \ref{sec:demixing}).
A direct comparison of the Uranus versus Neptune models can be found in Section \ref{sec:compare_uranus_neptune}.
We discuss limitations and the required follow-up work in Section \ref{sec:discussion} and present conclusions in Section \ref{sec:conclusions}.

%%%%%%%%%%%%%%%%%%%%%%%%%%%%%%%%%%%%%%%%%%%%%%%%%%%%%%%%%%%%%%%%%%%%%%%%%%%%%%%%%%%%%%%%%%
%%%%%%%%%%%%%%%%%%%%%%%%%%%%%%%%%%%%%%%%%%%%%%%%%%%%%%%%%%%%%%%%%%%%%%%%%%%%%%%%%%%%%%%%%%

\section{Methods}
\label{sec:methods}

%%%%%%%%%%%%%%%%%%%%%%%%%%%%%%%%%%%%%%%%%%%%%%%%%%%%%%%%%%%%%%%%%%%%%%%%%%%%%%%%%%%%%%%%%%

Interior models are designed to fit the measured planetary data. 
In this study we focused on Uranus and Neptune and used the data listed in Table \ref{tab:measured_data}. 

%%%%%%%%%%%%%%%%%%%%%%%%%%%%%%%%%%%%%%%%%%%%%%%%%%%%%%%%%%%%%%%%%%%%%%%%%%%%%%%%%%%%%%%%%%

\begin{figure}
    \centering
    \includegraphics[width=\hsize]{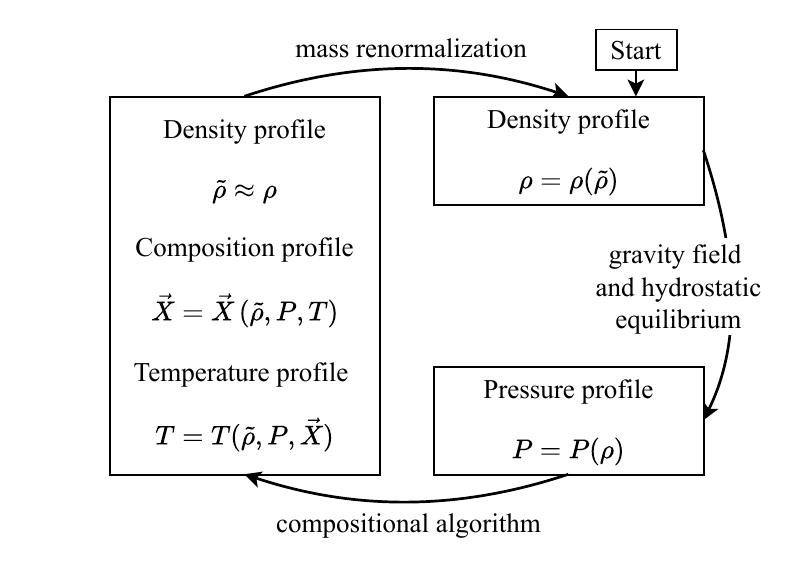}
    \caption{
    Overview of our global algorithm to infer agnostic and self-consistent planetary interior models. 
    The difference between $\tilde{\rho}$ used by the compositional algorithm and $\rho$ compatible with the interior pressure and planetary mass decreases until it vanishes using our iterative process.
    Traditionally, empirical models only use a single iteration, leading to inconsistencies. 
    Figure \ref{fig:global_algorithm} contains more details.
    }
    \label{fig:high_level_overview_algorithm}
\end{figure}

%%%%%%%%%%%%%%%%%%%%%%%%%%%%%%%%%%%%%%%%%%%%%%%%%%%%%%%%%%%%%%%%%%%%%%%%%%%%%%%%%%%%%%%%%%

We present a global algorithm to generate agnostic and physical interior models. 
A schematic overview of the algorithm is shown in Figure \ref{fig:high_level_overview_algorithm}. 
A more detailed description is presented in Figure \ref{fig:global_algorithm}. 
The global algorithm starts by generating an initial random density profile $\rho^{(0)}(r)$: 
We assume a perfectly spherical planet with radius $R$ and mass $M$. 
Spherical symmetry allows to describe the interior by a one-dimensional density function $\rho=\rho(r)$, where $r$ denotes the distance to the centre. 
Regarding the form of said function, every monotonically decreasing function $f: r\in[0,R] \rightarrow f(r)\in[1, 0]$ can be rescaled to a one-dimensional density profile $\rho$ via
\begin{equation}
    \rho(r) := \frac{f(r) \cdot M}{4\pi \int r^2 f(r) \mathrm{d}r}.
    \label{eq:renormalise_density}
\end{equation}
We generated a random initial function $f$ analogous to \cite{Podolak2022}: 
Start defining $f$ with the two tuples $(r=0,f(r)=1)$ and $(r=R, f(r)=0)$. 
These two tuples are interpreted as corners of the rectangle $[0,R] \times [1,0]$. 
Now choose a random point $(r^*, f(r^*))$ in the rectangle, adding a third tuple to the definition of $f$. 
The point $(r^*, f(r^*))$ defines two new and smaller rectangles: $[0,r^*] \times [1,f(r^*)]$ and $[r^*,R] \times [f(r^*),0]$. 
Within these smaller rectangles, we can again choose new tuples, increasing the number of points where $f$ is defined from three to five. 
Now repeat this procedure by choosing new tuples for each new rectangle as often as desired. 
This procedure ensures that $f$ is indeed monotonically decreasing and generated in a random manner. 
For this work, we chose $1+2+4+8+16=31$ tuples in addition to the two initial tuples and used linear interpolation between the tuples to get to $N=512$ equidistantly spaced radial points with a corresponding value of $f$. 
Applying a Gaussian filter to $f$ \citep[as in][]{Podolak2022} and using Equation \ref{eq:renormalise_density} yielded the initial random density $\rho^{(0)}(r)$.

%%%%%%%%%%%%%%%%%%%%%%%%%%%%%%%%%%%%%%%%%%%%%%%%%%%%%%%%%%%%%%%%%%%%%%%%%%%%%%%%%%%%%%%%%%

As a next step, $\rho^{(0)}(r)$ was given to the Theory of Figures\footnote{\url{https://doi.org/10.5281/zenodo.16902935} provides the numerical implementation.} \citep[][]{Zharkov1978, Morf2024}.
The Theory of Figures (ToF) takes the one-dimensional density function $\rho(r)$, the mass $M$, the equatorial radius $R_\text{eq}$, and the rotation period $P_\text{rot}$ as inputs and calculates the implied two-dimensional spheroidal shape of the planet. 
In particular, the gravitational moments $J_{n,\text{ToF}}$ and the internal pressure $P_\text{ToF}(r)$ are calculated as an output. 
The pressure $P_\text{ToF}(r)$ is calculated by integrating the hydrostatic equilibrium equation: 
\begin{equation}
    \vec{\nabla}P = \rho \vec{\nabla} U,
    \label{eq:HE}
\end{equation}
where $U$ is the total interior potential consisting of gravitational and centrifugal contributions \citep[Appendix A.1 in][for details]{Morf2024}.

%%%%%%%%%%%%%%%%%%%%%%%%%%%%%%%%%%%%%%%%%%%%%%%%%%%%%%%%%%%%%%%%%%%%%%%%%%%%%%%%%%%%%%%%%%

Having obtained initial density $\rho^{(0)}(r)$ and pressure $P_\text{ToF}^{(0)}(r)$ profiles, the compositional algorithm is called next. 
The details of the compositional algorithm are discussed in Appendix \ref{sec:compositional_algorithm}. 
In a nutshell, the compositional algorithm uses Equations of State (EoS) for different materials to infer a temperature $T_\text{EoS}^{(0)}(r)$ and composition profile $\vec{X}_\text{EoS}^{(0)}(r)$ based on the given density-pressure profile.
However, the densities implied by the EoS $\rho_\text{EoS}^{(0)}(r)$ can deviate from the originally provided densities $\rho^{(0)}(r)$. 
Although the compositional algorithm tries to keep the deviations as small as possible, they are necessary because a random density profile will almost never represent a physically realistic interior. 

%%%%%%%%%%%%%%%%%%%%%%%%%%%%%%%%%%%%%%%%%%%%%%%%%%%%%%%%%%%%%%%%%%%%%%%%%%%%%%%%%%%%%%%%%%

Due to these deviations, the density profile has to be renormalised (like in Equation \ref{eq:renormalise_density}) to ensure that the planet still has the correct mass $M$. 
The renormalised density profile is denoted as $\rho^{(1)}(r)$. 
Using the ToF a second time provides new values for the pressure $P_\text{ToF}^{(1)}(r)$ and gravitational moments $J^{(1)}_{n,\text{ToF}}$ that are consistent with $\rho^{(1)}(r)$. 
At this point, one hence has a self-consistent tuple: 
\begin{equation}
    \left(\rho^{(1)}(r), P_\text{ToF}^{(1)}(r), J^{(1)}_{n,\text{ToF}}\right), 
    \label{eq:rho_p_J_tuple}
\end{equation}
and an inferred temperature and composition tuple: 
\begin{equation}
    \left(T_\text{EoS}^{(0)}(r), \vec{X}_\text{EoS}^{(0)}(r)\right).
    \label{eq:T_comp_tuple}
\end{equation}
Combining Equations \ref{eq:rho_p_J_tuple} and \ref{eq:T_comp_tuple} to a full interior model is not self-consistent: 
The tuple in Equation \ref{eq:T_comp_tuple} is consistent with $\rho_\text{EoS}^{(0)}(r) \neq \rho^{(1)}(r)$ and $P_\text{ToF}^{(0)}(r) \neq P_\text{ToF}^{(1)}(r)$. 

%%%%%%%%%%%%%%%%%%%%%%%%%%%%%%%%%%%%%%%%%%%%%%%%%%%%%%%%%%%%%%%%%%%%%%%%%%%%%%%%%%%%%%%%%%

\begin{figure*}
    \centering
    \includegraphics[width=0.49\hsize]{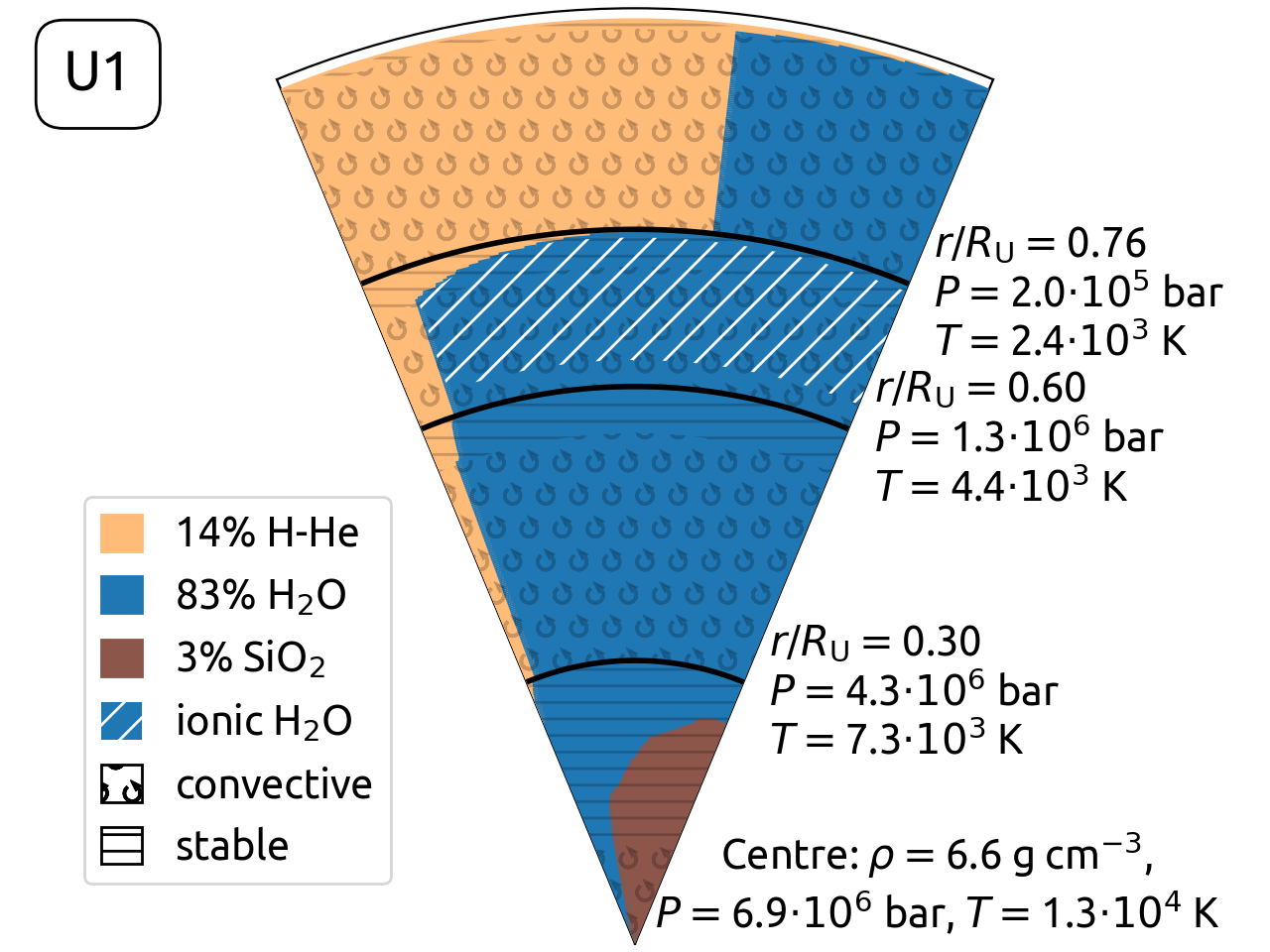}
    \includegraphics[width=0.49\hsize]{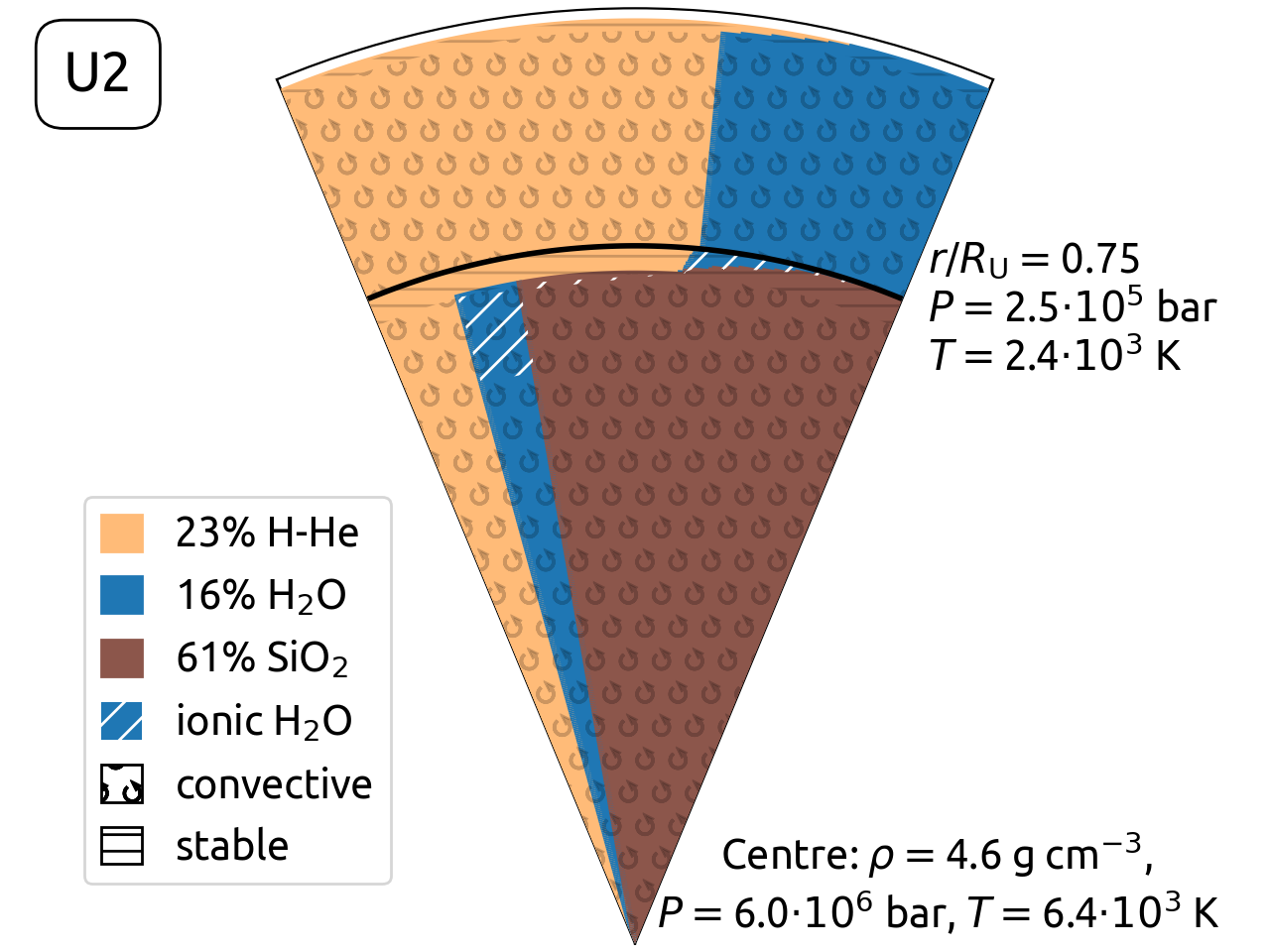}
    \includegraphics[width=0.49\hsize]{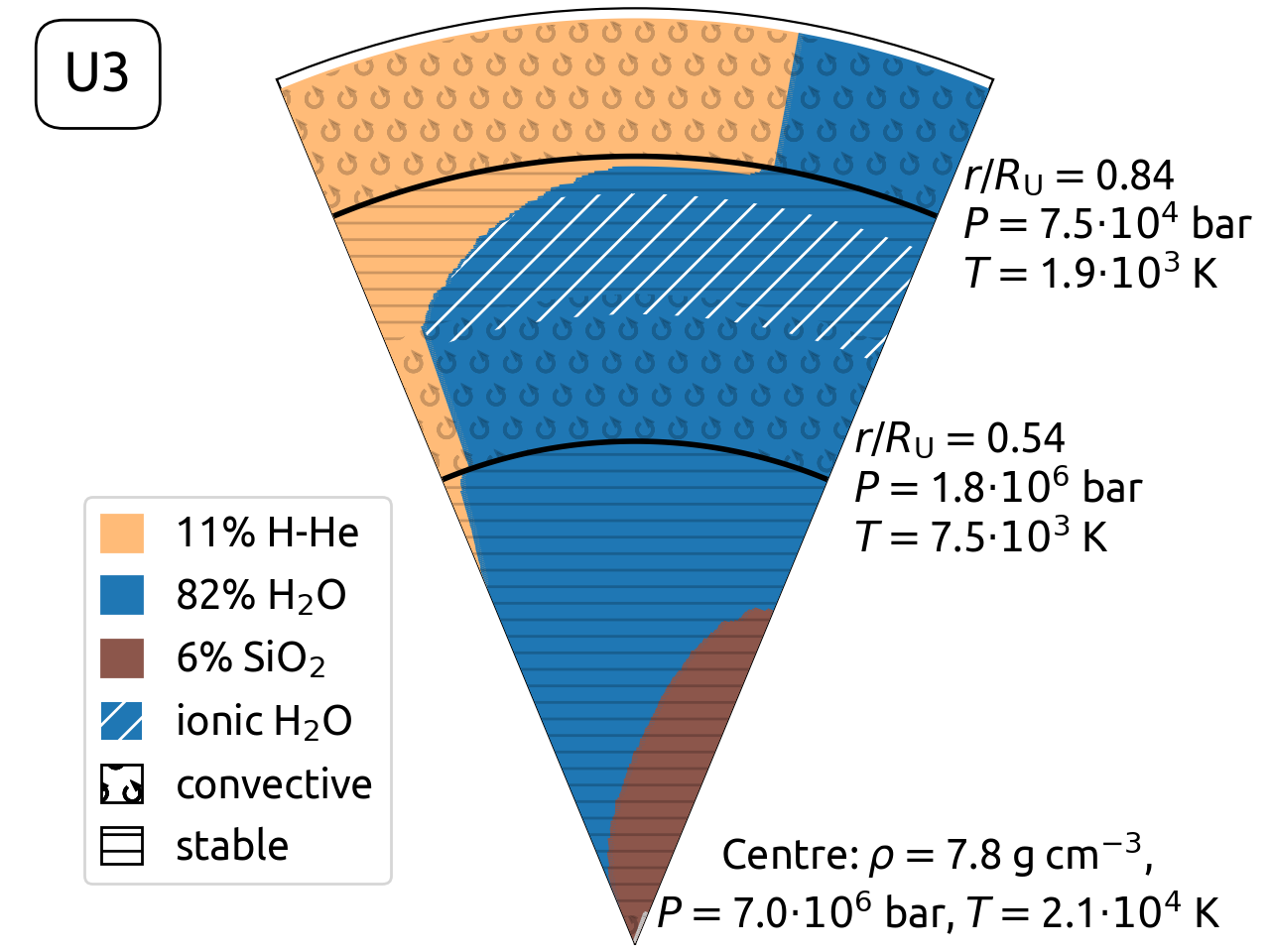}
    \includegraphics[width=0.49\hsize]{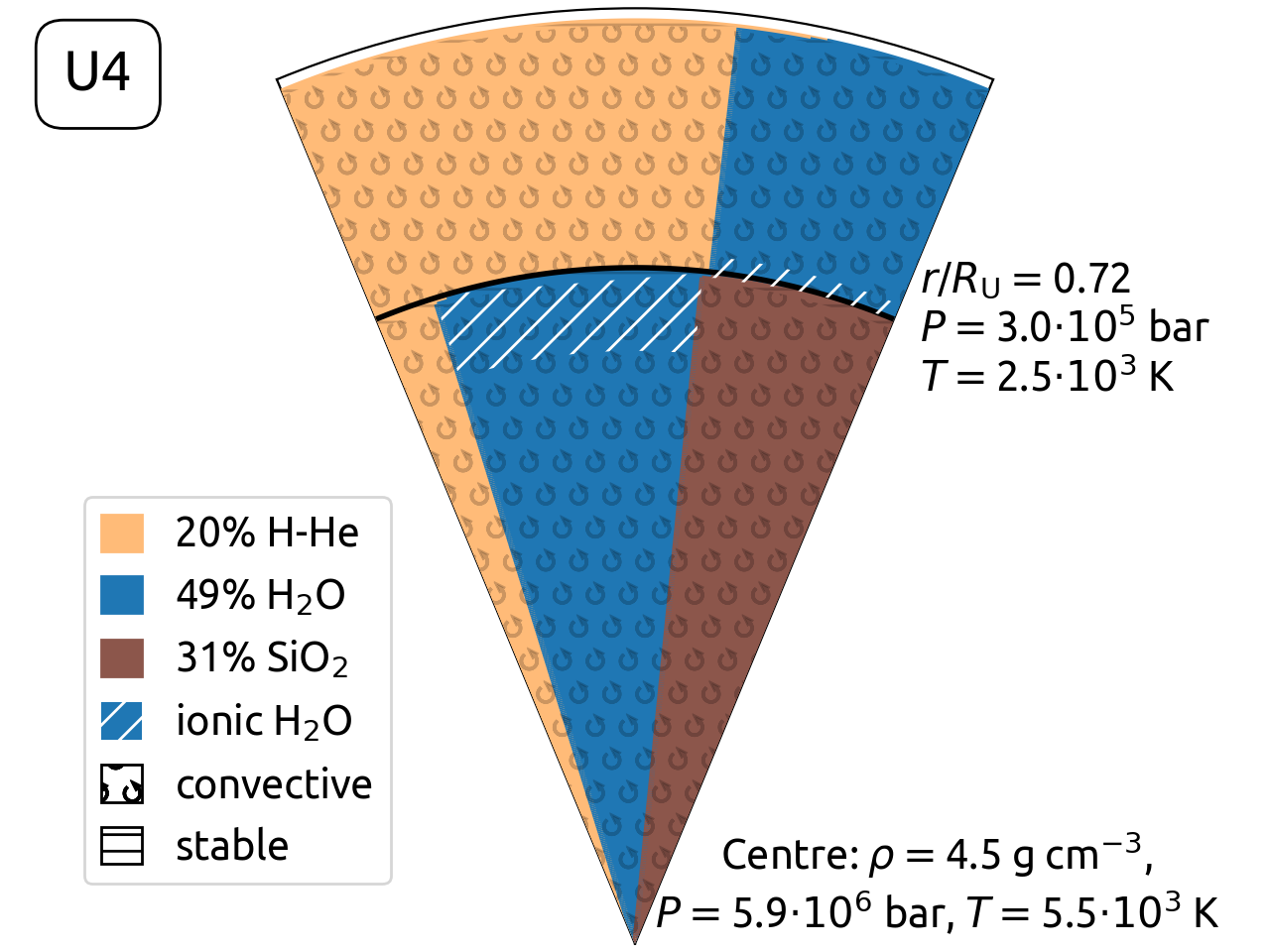}
    \caption{
    Inferred composition and convective and radiative structures for the four Uranus interior models.
    The legends summarise the total mass fractions for each component.
    The uppermost small arc of each slice is empty because we do not infer a composition for pressures below 100 bars.
    }
    \label{fig:Uranus_composition_slices}
\end{figure*}

%%%%%%%%%%%%%%%%%%%%%%%%%%%%%%%%%%%%%%%%%%%%%%%%%%%%%%%%%%%%%%%%%%%%%%%%%%%%%%%%%%%%%%%%%%

However, $\rho^{(1)}(r)$ should be much more compatible with the EoS than $\rho^{(0)}(r)$ was initially. 
By calling the compositional algorithm again, one can expect that it can more easily infer temperature $T_\text{EoS}^{(1)}(r)$ and composition $\vec{X}_\text{EoS}^{(1)}(r)$ profiles, whose implied densities $\rho_\text{EoS}^{(1)}(r)$ are close to the provided densities $\rho^{(1)}(r)$.
Consequently, $\rho^{(2)}(r)$ and $\rho^{(1)}(r)$ should be much closer to each other compared to how close $\rho^{(1)}(r)$ is to $\rho^{(0)}(r)$. 
Therefore, by repeating the above outlined procedure, eventually there should be a $k\in\mathbb{N}$, where the difference between $\rho^{(k+1)}(r)$ and $\rho^{(k)}(r)$ is so small for all $r$, that the tuple:  
\begin{equation}
    \left(\rho^{(k+1)}(r), P_\text{ToF}^{(k+1)}(r), J^{(k+1)}_{n,\text{ToF}}, T_\text{EoS}^{(k)}(r), \vec{X}_\text{EoS}^{(k)}(r)\right)
    \label{eq:complete_tuple}
\end{equation}
is practically self-consistent. 
Then, Equation \ref{eq:complete_tuple} is a complete physical model of a planetary interior which has been generated in a random manner. 
We employed the criterion: 
\begin{equation}
    \text{for all } r: \frac{\left| P_\text{ToF}^{(k+1)}(r) - P_\text{ToF}^{(k)}(r) \right|}{P_\text{ToF}^{(k+1)}(r)} < \epsilon 
    \label{eq:pressure_conistency}
\end{equation}
to check whether a solution is self-consistent. 
Since the density and pressure are directly linked to each other via Equation \ref{eq:HE}, there is no need to define a second criterion for the density. 
In this work, we chose $\epsilon = 0.02$. 
This is a conservative number given the uncertainties associated with the EoS of the different materials and their mixtures.  
For example, the assumption of ideal mixing (Equation \ref{eq:ideal_mixing}) alone can introduce deviations of up to 5\% in density \citep[][]{Darafeyeu2024}.
A detailed discussion about algorithm convergence is given in Appendix \ref{sec:algorithm_convergence}.

%%%%%%%%%%%%%%%%%%%%%%%%%%%%%%%%%%%%%%%%%%%%%%%%%%%%%%%%%%%%%%%%%%%%%%%%%%%%%%%%%%%%%%%%%%

Finally, in general $J^{(k+1)}_{n,\text{ToF}} \neq J^{(0)}_{n,\text{ToF}}$ holds true at the end of the global algorithm. 
So there is no point in ensuring a precise match between the initial gravitational moments $J^{(0)}_{n,\text{ToF}}$ and the available data in Table \ref{tab:measured_data}. 
Only $J^{(k+1)}_{n,\text{ToF}}$ has to be data consistent, promoting Equation \ref{eq:complete_tuple} from a physical interior model to a model also compatible with the data of the considered planet. 

%%%%%%%%%%%%%%%%%%%%%%%%%%%%%%%%%%%%%%%%%%%%%%%%%%%%%%%%%%%%%%%%%%%%%%%%%%%%%%%%%%%%%%%%%%

To summarise, we iterated until the density, pressure, temperature, and composition profile have converged and all agreed with the data in Table \ref{tab:measured_data} and the used EoS. 
Our method is capable of generating interior models that are both agnostic and self-consistent. 
Being agnostic refers to using a minimal set of assumptions like for empirical models.
However, we still maintain a level of self-consistency typical for physical models.
As shown below, our results highlight that no single interior model can be regarded as the standard for a given planet. 
A wide range of self-consistent models must be considered to get a full picture that can reliably inform formation and evolution models. 

%%%%%%%%%%%%%%%%%%%%%%%%%%%%%%%%%%%%%%%%%%%%%%%%%%%%%%%%%%%%%%%%%%%%%%%%%%%%%%%%%%%%%%%%%%
%%%%%%%%%%%%%%%%%%%%%%%%%%%%%%%%%%%%%%%%%%%%%%%%%%%%%%%%%%%%%%%%%%%%%%%%%%%%%%%%%%%%%%%%%%

\section{Uranus}
\label{sec:Uranus}

%%%%%%%%%%%%%%%%%%%%%%%%%%%%%%%%%%%%%%%%%%%%%%%%%%%%%%%%%%%%%%%%%%%%%%%%%%%%%%%%%%%%%%%%%%

\begin{figure}
    \centering
    \includegraphics[width=\hsize]{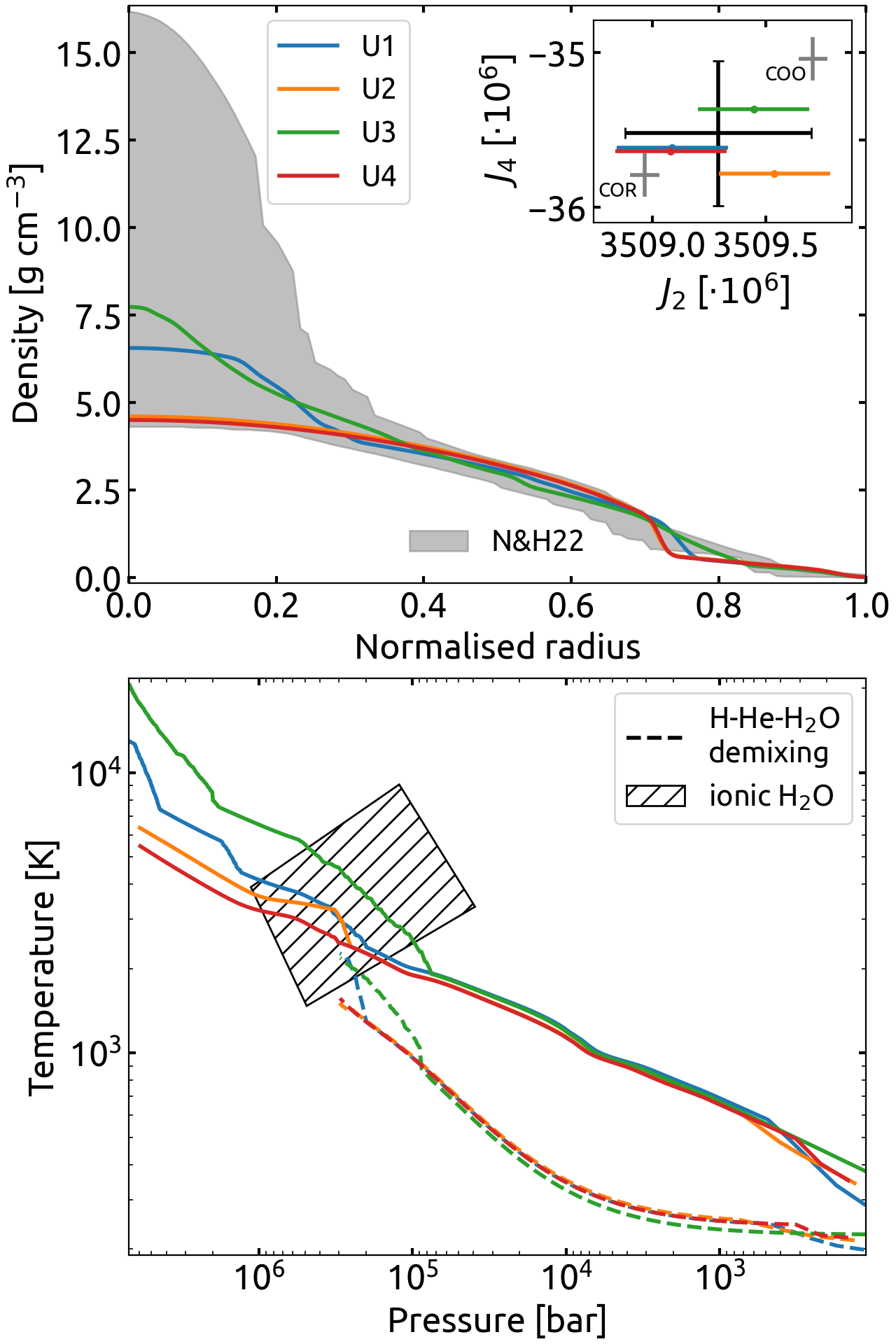}
    \caption{
    Density, pressure, and temperature profiles for the four Uranus models.
    {\bf Top:} Density versus normalised radius. 
    For comparison, also shown are empirical solutions from \cite{Neuenschwander2022} (grey area).
    The panel in the top right shows  Uranus' measured \citep[][black]{French2024} and calculated (coloured) gravitational moments.
    We also show Uranus' gravitational moments according to \cite{Jacobson2025} (grey) who find different results (COR vs COO) corresponding to different ring centre definitions.
    The coloured uncertainties depict estimated errors from the Theory of Figures.
    {\bf Bottom:} Temperature versus pressure. 
    Also shown is the predicted region of ionic water inferred by \cite{Redmer2011}.
    The dashed lines show the demixing line, below which hydrogen, helium, and water are expected to become immiscible \citep{Howard2025}.
    }
    \label{fig:Uranus_densities_TvsP}
\end{figure}

%%%%%%%%%%%%%%%%%%%%%%%%%%%%%%%%%%%%%%%%%%%%%%%%%%%%%%%%%%%%%%%%%%%%%%%%%%%%%%%%%%%%%%%%%%

We present four Uranus models: U1, U2, U3, and U4.
The inferred internal structures and compositions of the Uranus models are summarised in Figure \ref{fig:Uranus_composition_slices}. 
Models U1 and U3 correspond to water-rich interior solutions.  
Their respective rock-to-water ratios are low, just 0.04 and 0.08, consistent with Uranus being an 'ice giant'.
These models also possess similar total H-He abundances (14\% and 11\% of Uranus' mass, respectively).
However, their convection and temperature profiles vary significantly.
Model U1 is mostly convective, has four different convection zones and reaches a central temperature below 13,000 K.
On the other hand, model U3 has two large stable regions.
The innermost stable region covers more than half (54\%) of Uranus's radius and the central temperature is above 20,000 K. 

%%%%%%%%%%%%%%%%%%%%%%%%%%%%%%%%%%%%%%%%%%%%%%%%%%%%%%%%%%%%%%%%%%%%%%%%%%%%%%%%%%%%%%%%%%

On the other hand, models U2 and U4 are very similar in terms of their convection and thermal profiles.
Both models reach a central temperature of $\sim$ 6,000 K.
They are almost exclusively convective with two large convection zones separated by a small stable layer at $\sim$ 75\% of Uranus' radius.
However, the inferred compositions in models U2 and U4 are very different. 
In model U2, Uranus' interior is rock-dominated with a rock-to-water ratio of 3.92, while the rock-to-water ratio in model U4 is 0.64. 
The H-He abundances are more comparable: 
23\% and 20\% of Uranus' mass for models U2 and U4, respectively. 
In addition, H-He is present also in the planetary deep interior in both cases.  
The H-He mass fractions for models U2 and U4 in the centre are 16\% and 12\%, respectively. 

%%%%%%%%%%%%%%%%%%%%%%%%%%%%%%%%%%%%%%%%%%%%%%%%%%%%%%%%%%%%%%%%%%%%%%%%%%%%%%%%%%%%%%%%%%

The density profiles and temperature-pressure profiles are presented in Figure \ref{fig:Uranus_densities_TvsP}.
We find that the density profiles for models U2 and U4 are very similar. 
They both have a density discontinuity ($\Delta\rho>$ 1 g cm$^{-3}$) at $\sim$ 75\% of Uranus' radius and reach central densities of $\sim$4.5 g cm$^{-3}$.
This density discontinuity around $3\cdot10^5$ bar leads to a temperature jump that is 500 K higher for U2 compared to U4.
Models U1 and U3 reach higher central densities (6.6 and 7.8 g cm$^{-3}$, respectively) than U2 and U4.
There are also no clear jumps in density in models U1 and U3.
We further observe that all models leave the solution space discovered by \cite{Neuenschwander2022} around a normalised radius of 0.8.
This finding demonstrates how assumptions made by the modeller (such as the use of three consecutive polytropes in \cite{Neuenschwander2022}) can result in biased outcomes and exclude parts of the solution space.
In contrast, our randomised approach represents a notable improvement, expanding the scope of the solution space.

%%%%%%%%%%%%%%%%%%%%%%%%%%%%%%%%%%%%%%%%%%%%%%%%%%%%%%%%%%%%%%%%%%%%%%%%%%%%%%%%%%%%%%%%%%

For comparison, we show the measured gravitational moments from \cite{Jacobson2025} (in grey) alongside those from \cite{French2024} (in black) in Figure \ref{fig:Uranus_densities_TvsP}.
\cite{Jacobson2025} only report formal (not realistic) uncertainties for their two results based on precession rates of either the ring centres of opacity (COO) or the geometric ring midlines (COR).
They conclude that their knowledge of the gravitational moments is not better compared to that postulated by \cite{French2024}.

%%%%%%%%%%%%%%%%%%%%%%%%%%%%%%%%%%%%%%%%%%%%%%%%%%%%%%%%%%%%%%%%%%%%%%%%%%%%%%%%%%%%%%%%%%

Overall, our results show that both rock-dominated and water-dominated solutions are possible for Uranus.
The same holds true for further properties:
A wide variety of density, temperature, and convection profiles are feasible.
The temperature and entropy profiles of our Uranus models can be found in Appendix \ref{sec:additional_data} in Figure \ref{fig:physical_solutions_temperatures_entropies_N2_gen_4_U}.
Relative pressure deviations according to Equation \ref{eq:pressure_conistency} are shown in Figure \ref{fig:physical_solutions_pressure_deltas_gen_4_U}.

%%%%%%%%%%%%%%%%%%%%%%%%%%%%%%%%%%%%%%%%%%%%%%%%%%%%%%%%%%%%%%%%%%%%%%%%%%%%%%%%%%%%%%%%%%
%%%%%%%%%%%%%%%%%%%%%%%%%%%%%%%%%%%%%%%%%%%%%%%%%%%%%%%%%%%%%%%%%%%%%%%%%%%%%%%%%%%%%%%%%%

\section{Neptune}
\label{sec:Neptune}

%%%%%%%%%%%%%%%%%%%%%%%%%%%%%%%%%%%%%%%%%%%%%%%%%%%%%%%%%%%%%%%%%%%%%%%%%%%%%%%%%%%%%%%%%%

\begin{figure*}
    \centering
    \includegraphics[width=0.49\hsize]{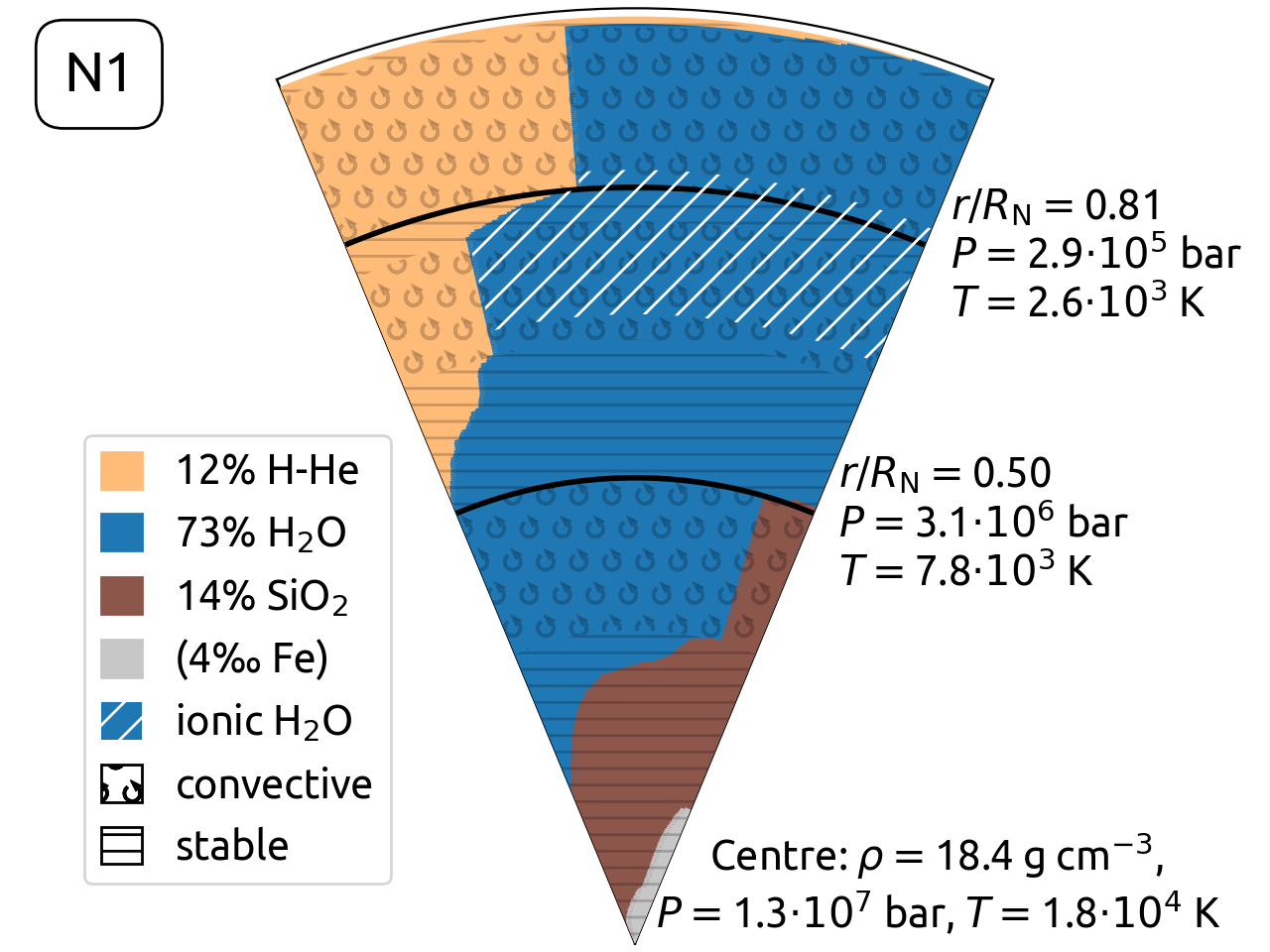}
    \includegraphics[width=0.49\hsize]{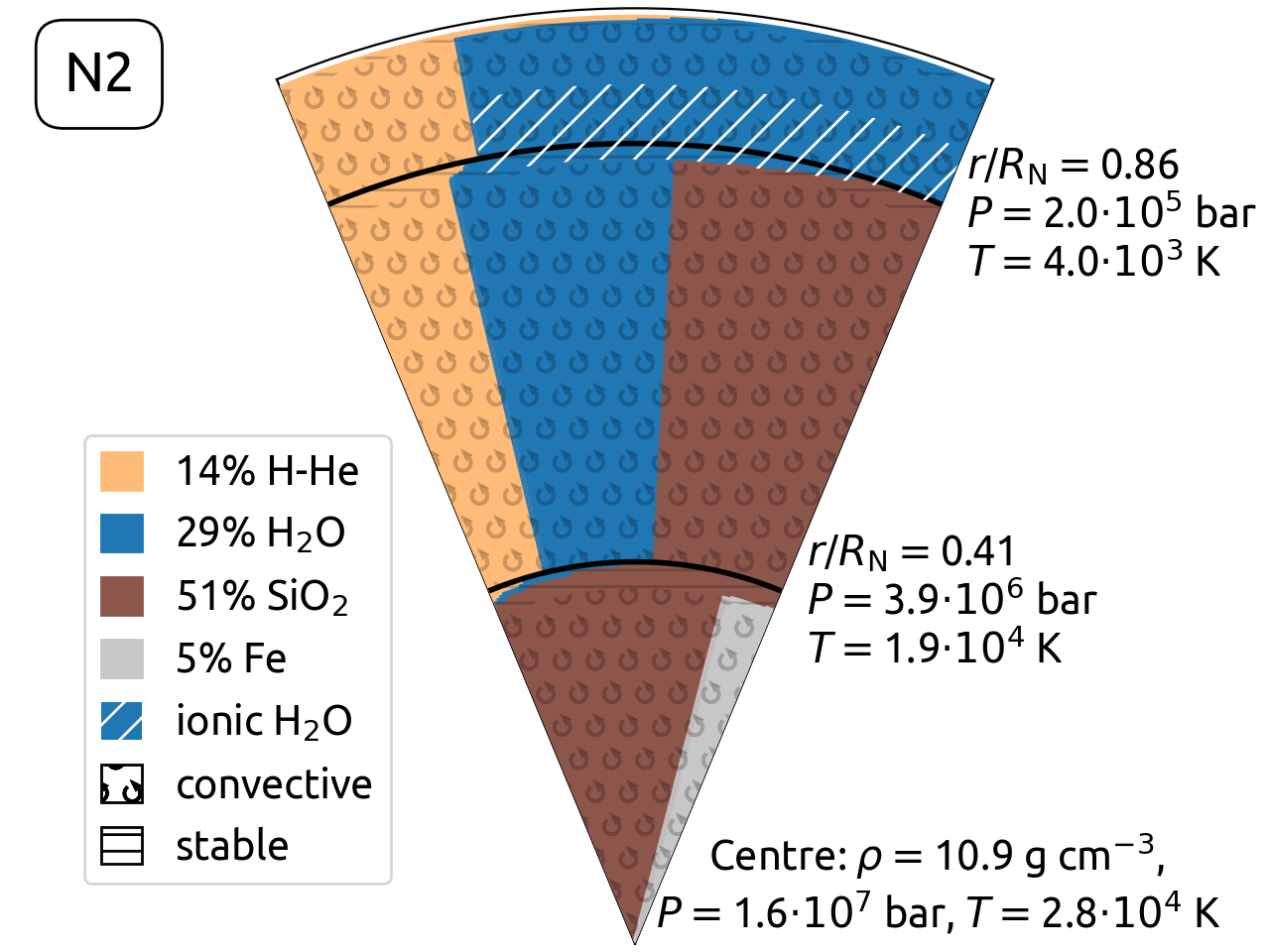}
    \includegraphics[width=0.49\hsize]{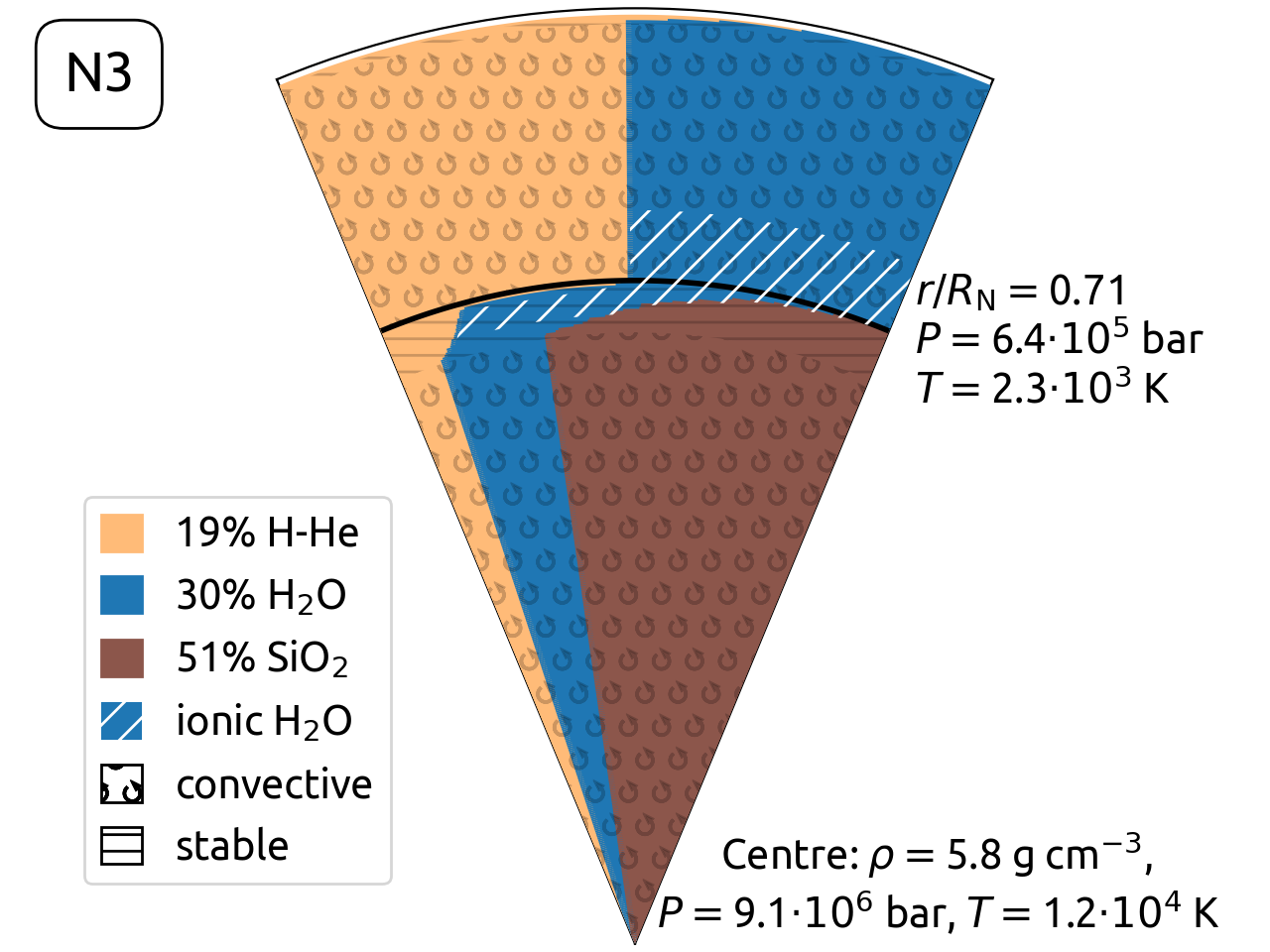}
    \includegraphics[width=0.49\hsize]{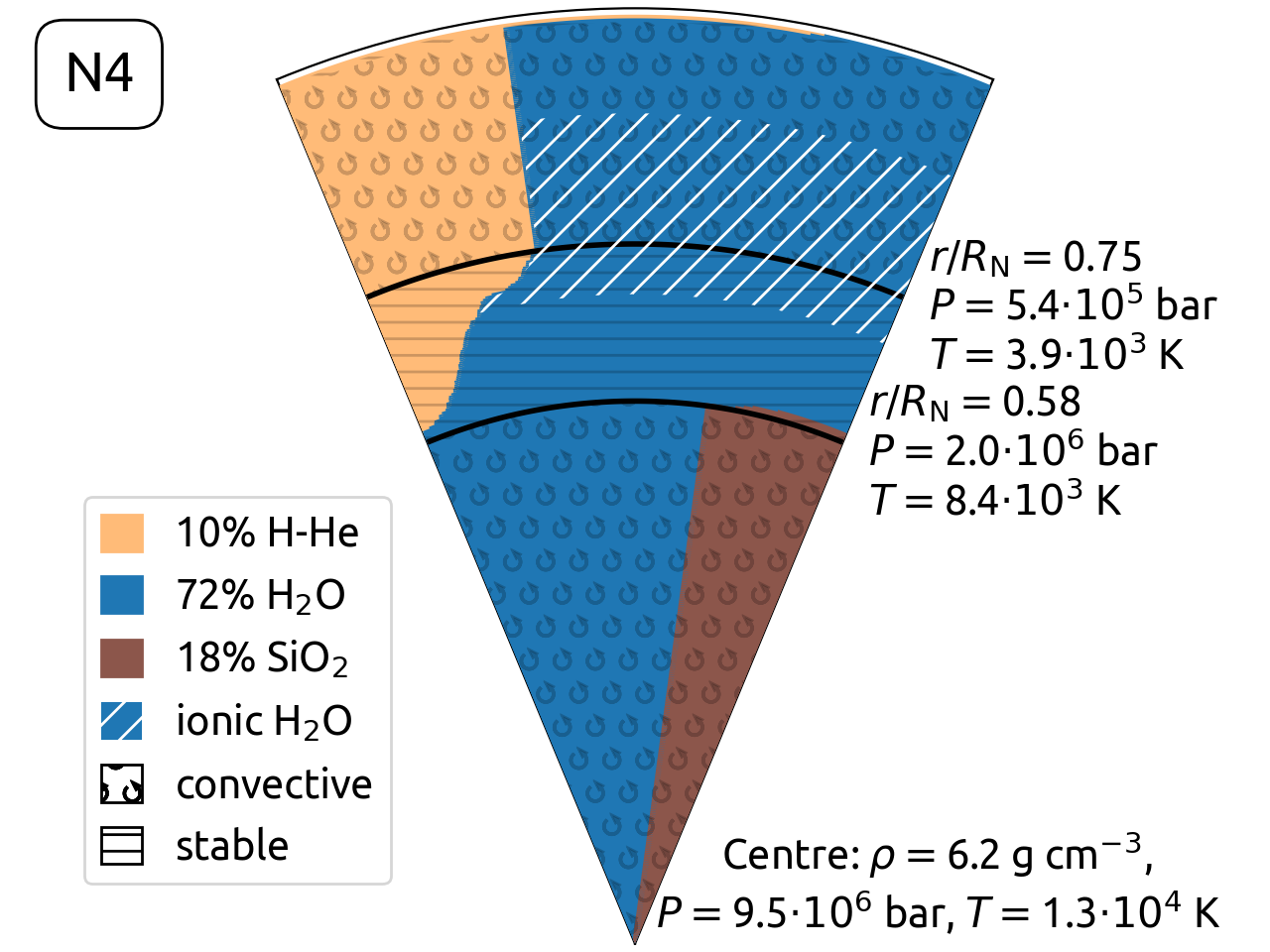}
    \caption{
    Same as Figure \ref{fig:Uranus_composition_slices}, but for Neptune.
    }
    \label{fig:Neptune_composition_slices}
\end{figure*}

%%%%%%%%%%%%%%%%%%%%%%%%%%%%%%%%%%%%%%%%%%%%%%%%%%%%%%%%%%%%%%%%%%%%%%%%%%%%%%%%%%%%%%%%%%

\begin{figure}
    \centering
    \includegraphics[width=\hsize]{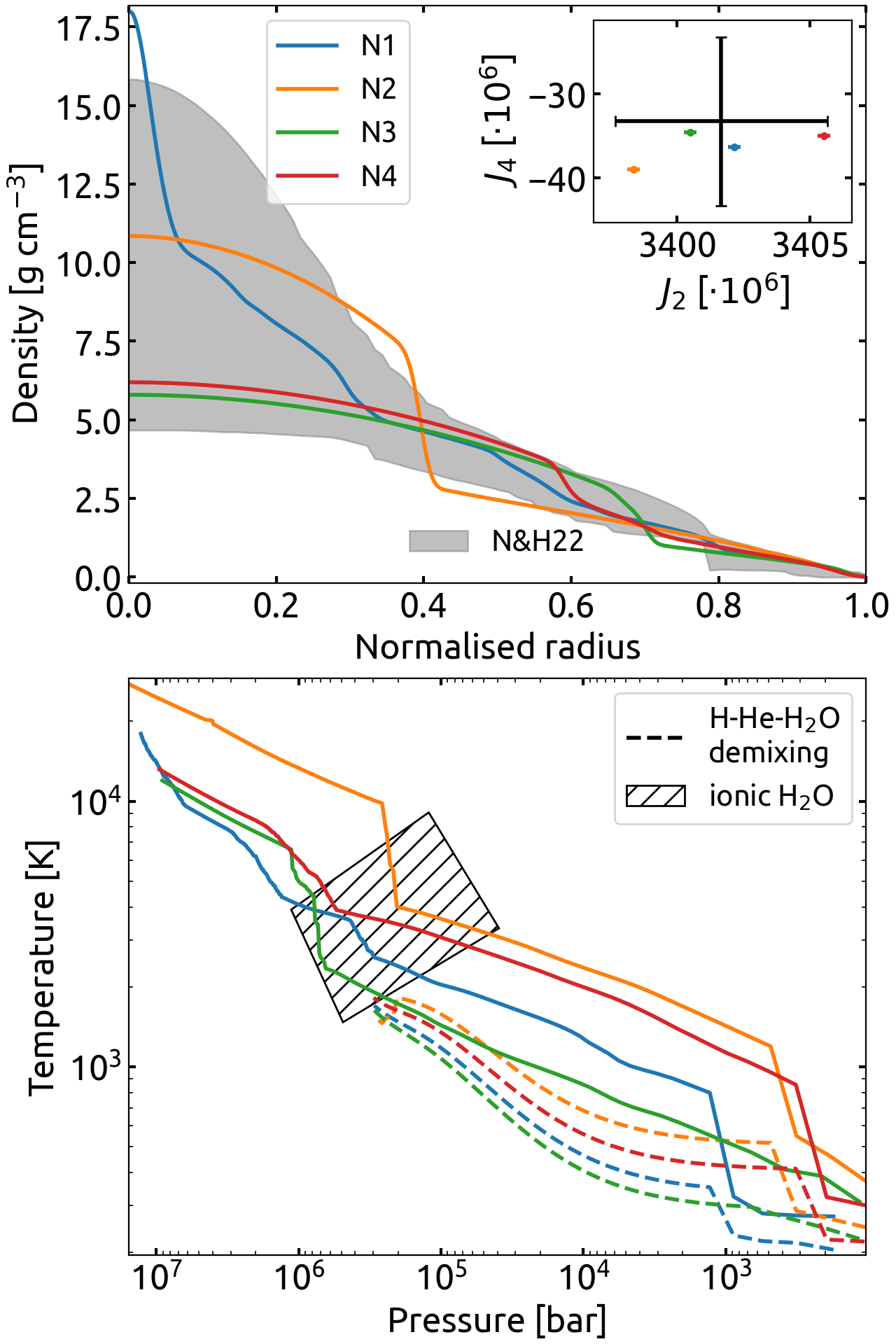}
    \caption{
    Same as Figure \ref{fig:Uranus_densities_TvsP}, but for Neptune.
    }
    \label{fig:Neptune_densities_TvsP}
\end{figure}

%%%%%%%%%%%%%%%%%%%%%%%%%%%%%%%%%%%%%%%%%%%%%%%%%%%%%%%%%%%%%%%%%%%%%%%%%%%%%%%%%%%%%%%%%%

We present four Neptune models N1, N2, N3 and N4. 
Figure \ref{fig:Neptune_composition_slices} shows sketches of the inferred structure and bulk composition of our Neptune models. 
Models N1 and N4 are consistent with the notion of Neptune being an 'ice giant'.
They have water-dominated interiors with rock-to-water ratios of 0.20 and 0.26, respectively, and H-He mass fractions of 12\% and 10\%, respectively.
However, there are also significant differences.
Model N1 has a central density higher than 18 g cm$^{-3}$, while the central density of model N4 is only 6.2 g cm$^{-3}$.
Although both models include stable regions, they are more prevalent for model N1. 
For example, model N4 has a fully convective deep interior while the innermost region of model N1 is stable against convection and has a composition gradient.
Model N4 initially has a larger temperature gradient compared to model N1, leading to a first appearance of ionic water at relatively larger normalised radii (0.89 vs 0.83) with higher temperatures (3100 K vs 2500 K).

%%%%%%%%%%%%%%%%%%%%%%%%%%%%%%%%%%%%%%%%%%%%%%%%%%%%%%%%%%%%%%%%%%%%%%%%%%%%%%%%%%%%%%%%%%

Models N2 and N3 represent Neptune models that are rock-dominated, supporting the idea of Neptune being a 'rock giant' \cite[for example][]{Helled2020, Teanby2020}.
Their rock-to-water ratios are found to be 1.78 and 1.72, with more than 50\% of the planetary mass consisting of rocks.
Although both models have large convective zones, the interior temperature profiles are rather different. 
Model N2 has a central temperature of 28,000 K, while the central temperature of model N3 is 12,000 K.
Both of these inferred temperatures are significantly higher than the temperature inferred by standard adiabatic models which are of the order of 6000 K \citep[for example][]{Nettelmann2013}. 
In addition, in model N3 H-He is found also in the planetary centre with 10 \% of the composition by mass.
The existence of H-He in the deep interior is consistent with formation models of Neptune with composition gradients \citep[for example][]{Valletta2022}.
The central region in model N2, on the other hand, is enriched with iron (18 \% by mass).

%%%%%%%%%%%%%%%%%%%%%%%%%%%%%%%%%%%%%%%%%%%%%%%%%%%%%%%%%%%%%%%%%%%%%%%%%%%%%%%%%%%%%%%%%%

The inferred density profiles and pressure-temperature profiles are shown in Figure \ref{fig:Neptune_densities_TvsP}.
For Neptune, we find four very unique density profiles. 
Model N1 has the highest central density of 18.4 g cm$^{-3}$.
The density in model N2 is below 3 g cm$^{-3}$ in the outer half of the planet until it reaches a large density discontinuity ($\Delta\rho>3$ g cm$^{-3}$).
Models N3 and N4 have similar central densities (5.8 and 6.2 g cm$^{-3}$, respectively), but differ with respect to the location of their density discontinuity.
The discontinuities are located at a normalised radius of $\sim$0.7 for model N3 and $\sim$0.6 for model N2.
We further find that models N1, N2, and N4 have a temperature jump larger than 500 K between pressures of 200 and 2000 bars. 
Also in the case of Neptune, there are significant differences in the density profiles in comparison to the solution space found by \cite{Neuenschwander2022}.
This is observed at multiple locations (normalised radii of around 0.05, 0.4, and 0.75) with the differences being more notable compared to Uranus.
Our randomised approach indeed covers a wider space of solutions compared to parametrised approaches.

%%%%%%%%%%%%%%%%%%%%%%%%%%%%%%%%%%%%%%%%%%%%%%%%%%%%%%%%%%%%%%%%%%%%%%%%%%%%%%%%%%%%%%%%%%

We conclude that, like for Uranus, both rock-dominated and water-dominated solutions are valid, and that a wide range of compositions and internal structures is consistent with Neptune's observed data.
Our models consist of large convective zones or several stable regions, and can differ by more than 10000 K with respect to their core temperatures.
The temperature and entropy profiles of the four Neptune models can be found in Appendix \ref{sec:additional_data} in Figure \ref{fig:physical_solutions_temperatures_entropies_N2_gen_4_N}.
Relative pressure deviations according to Equation \ref{eq:pressure_conistency} are shown in Figure \ref{fig:physical_solutions_pressure_deltas_gen_4_N}.

%%%%%%%%%%%%%%%%%%%%%%%%%%%%%%%%%%%%%%%%%%%%%%%%%%%%%%%%%%%%%%%%%%%%%%%%%%%%%%%%%%%%%%%%%%

\begin{table}
\caption{
Total H-He mass fractions derived in this study (and previous work). 
}
\centering
\begin{tabular}{ll}
\hline
\hline
\multicolumn{2}{c}{Uranus} \\
\hline
This study & 0.11, 0.14, 0.20, 0.23 \\
\cite{Nettelmann2013} & 0.13, 0.15 \\
\cite{Militzer2024} & 0.02 \\
\cite{Neuenschwander2024} & 0.01 -- 0.20 \\
\cite{Malamud2024} & $\lesssim$ 0.25  \\
\cite{Morf2024} & $\lesssim$ 0.2 \\
\hline
\multicolumn{2}{c}{Neptune} \\
\hline
This study & 0.10, 0.12, 0.14, 0.19 \\
\cite{Nettelmann2013} & 0.16, 0.16, 0.19 \\
\cite{Militzer2024} & 0.04 \\
\cite{Malamud2024} & $\lesssim$ 0.15 \\
\hline
\hline
\end{tabular}
\label{tab:hydrogen_helium}
\end{table}

%%%%%%%%%%%%%%%%%%%%%%%%%%%%%%%%%%%%%%%%%%%%%%%%%%%%%%%%%%%%%%%%%%%%%%%%%%%%%%%%%%%%%%%%%%

\begin{figure}
    \centering
    \includegraphics[width=\hsize]{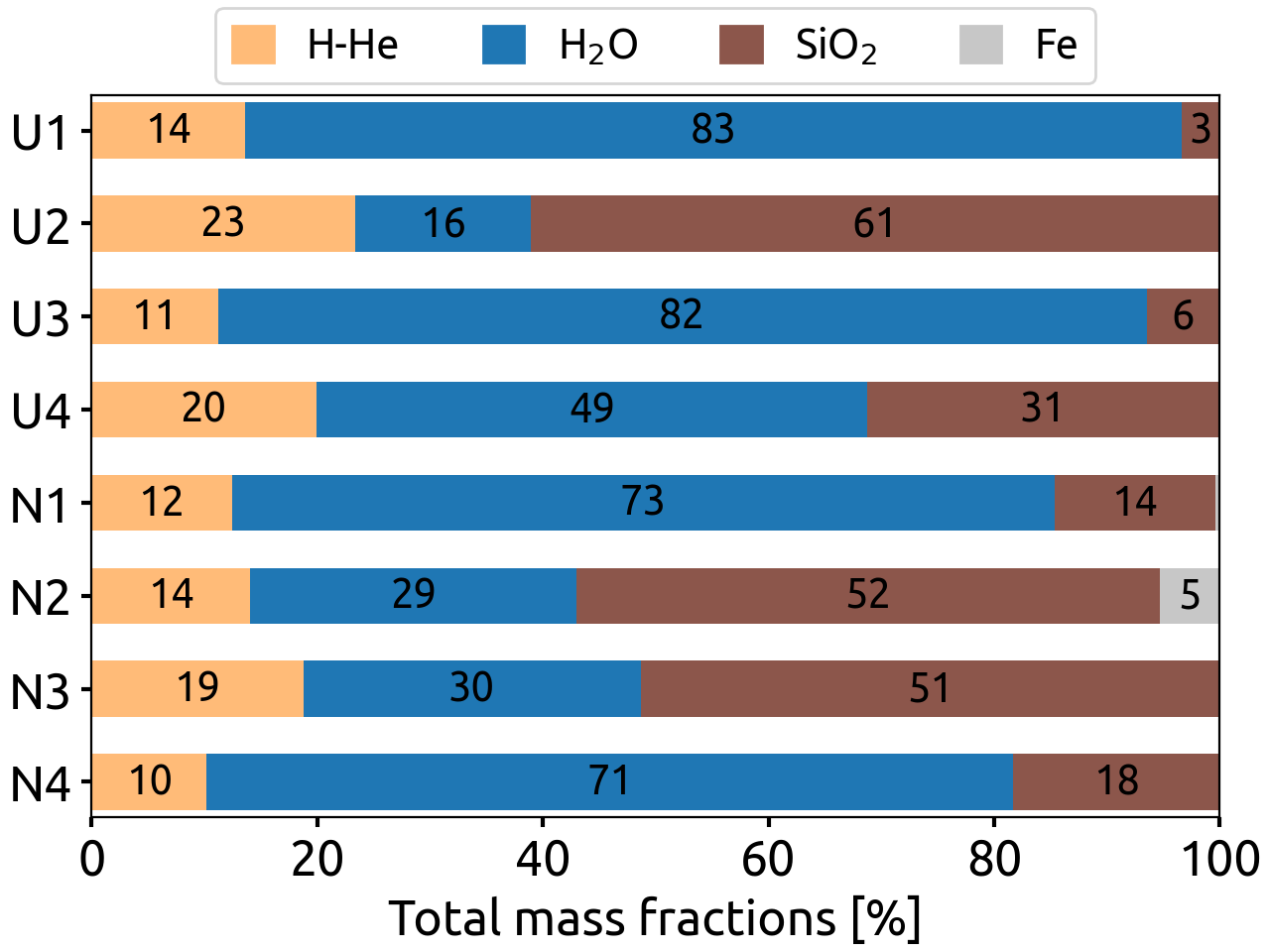}
    \caption{
    Inferred bulk composition for Uranus and Neptune. 
    }
    \label{fig:bulk_compositions}
\end{figure}

%%%%%%%%%%%%%%%%%%%%%%%%%%%%%%%%%%%%%%%%%%%%%%%%%%%%%%%%%%%%%%%%%%%%%%%%%%%%%%%%%%%%%%%%%%

\begin{table}
\caption{
Rock-to-water ratios derived in this study (and previous work). 
}
\centering
\begin{tabular}{ll}
\hline
\hline
\multicolumn{2}{c}{Uranus} \\
\hline
This study & 0.04, 0.08, 0.64, 3.92 \\
\cite{Nettelmann2013} & 0.03, 0.05 \\
\cite{Militzer2024} & 0.04 \\
\cite{Arevalo2025} & 0.67 \\
\cite{Neuenschwander2024} & 0.05 -- 0.38 \\
\cite{Malamud2024} incl. CH$_4$ & $\sim 10^{-1}$ -- $10^{1}$  \\
\cite{Morf2024} & $\lesssim$ $10^{1}$ \\
\hline
\multicolumn{2}{c}{Neptune} \\
\hline
This study & 0.20, 0.26, 1.72, 1.78 \\
\cite{Nettelmann2013} & 0.07, 0.27, 0.28 \\
\cite{Militzer2024} & 0.14 \\
\cite{Malamud2024} incl. CH$_4$ & $\sim 10^{-1}$ -- $10^{1}$ \\
\hline
\hline
\end{tabular}
\tablefoot{The results from \cite{Malamud2024} are taken from Appendix C and represent a (water+methane)-to-(rock+iron) ratio.}
\label{tab:rock_to_water}
\end{table}

%%%%%%%%%%%%%%%%%%%%%%%%%%%%%%%%%%%%%%%%%%%%%%%%%%%%%%%%%%%%%%%%%%%%%%%%%%%%%%%%%%%%%%%%%%
%%%%%%%%%%%%%%%%%%%%%%%%%%%%%%%%%%%%%%%%%%%%%%%%%%%%%%%%%%%%%%%%%%%%%%%%%%%%%%%%%%%%%%%%%%

\section{Implications} 
\label{sec:implications}

%%%%%%%%%%%%%%%%%%%%%%%%%%%%%%%%%%%%%%%%%%%%%%%%%%%%%%%%%%%%%%%%%%%%%%%%%%%%%%%%%%%%%%%%%%

\subsection{Bulk composition}
\label{sec:bulk_composition}

%%%%%%%%%%%%%%%%%%%%%%%%%%%%%%%%%%%%%%%%%%%%%%%%%%%%%%%%%%%%%%%%%%%%%%%%%%%%%%%%%%%%%%%%%%

The inferred bulk compositions of our Uranus and Neptune models are summarised in Figure \ref{fig:bulk_compositions}. 
We also list the total H-He mass fractions and compare them to previous work in Table \ref{tab:hydrogen_helium}. 
We find that the total H-He mass fractions (0.11--0.23 for Uranus and 0.10--0.19 for Neptune) are consistent with previous studies with the exception of \cite{Militzer2024}, who obtains lower values.
The derived rock-to-water ratios of our models (and selected previous studies) are listed in Table \ref{tab:rock_to_water}. 
We find that for both Uranus and Neptune, the rock-to-water ratio is poorly constrained and ranges between 0.04 and 3.92 for Uranus, and between 0.20 and 1.78 for Neptune. 
Overall, our models include solutions with rock-to-water ratios that are significantly higher compared to previous studies (up to 3.92 for Uranus and 1.78 for Neptune).
At the same time, we show that solutions with low rock-to-water ratios (as low as 0.04 for Uranus and 0.20 for Neptune) are also valid. 
We can therefore conclude that given the current data, there is no clear reason to prefer water-rich or rock-rich interiors for Uranus and Neptune. 
Therefore, referring to Uranus and Neptune as 'ice giants' may be inappropriate since the bulk composition of the planets remain unclear \citep[for example][]{Helled2020, Hofstadter2024}.
Constraining the rock-to-water ratio would require additional (and more accurate)  measurements \citep[for example][]{Nimmo2024} or additional constraints from  planetary formation models \citep[for example][]{Helled2025}.

%%%%%%%%%%%%%%%%%%%%%%%%%%%%%%%%%%%%%%%%%%%%%%%%%%%%%%%%%%%%%%%%%%%%%%%%%%%%%%%%%%%%%%%%%%

\subsection{Magnetic field generation}
\label{sec:magnetic_fields}

%%%%%%%%%%%%%%%%%%%%%%%%%%%%%%%%%%%%%%%%%%%%%%%%%%%%%%%%%%%%%%%%%%%%%%%%%%%%%%%%%%%%%%%%%%

Thanks to Voyager 2, we know that both Uranus and Neptune possess considerable magnetic fields \citep[][]{Connerney1987, Ness1989}.
The fields are multipolar and tilted significantly relative to the rotation axes of the planets \citep[for example][]{Jones2011}. 
These observations are relevant when modelling the planetary interiors, as solutions have to be consistent with dynamo generation. 
A common explanation for the magnetic fields of the planets is the presence of ionic water in a convective region \citep{Redmer2011}.
For example, \cite{Stanley2004, Stanley2006} found that a thin convective and conductive shell in the outer planetary region can suppress large-scale dipolar modes and hence explain the observed multipolar fields.
This occurs for both liquid (+stably stratified) and solid (+insulating) cores.
And \cite{Gastine2012} showed that multipolar fields can be reproduced for an even wider range of scenarios.

%%%%%%%%%%%%%%%%%%%%%%%%%%%%%%%%%%%%%%%%%%%%%%%%%%%%%%%%%%%%%%%%%%%%%%%%%%%%%%%%%%%%%%%%%%

\begin{figure}
    \centering
    \includegraphics[width=\hsize]{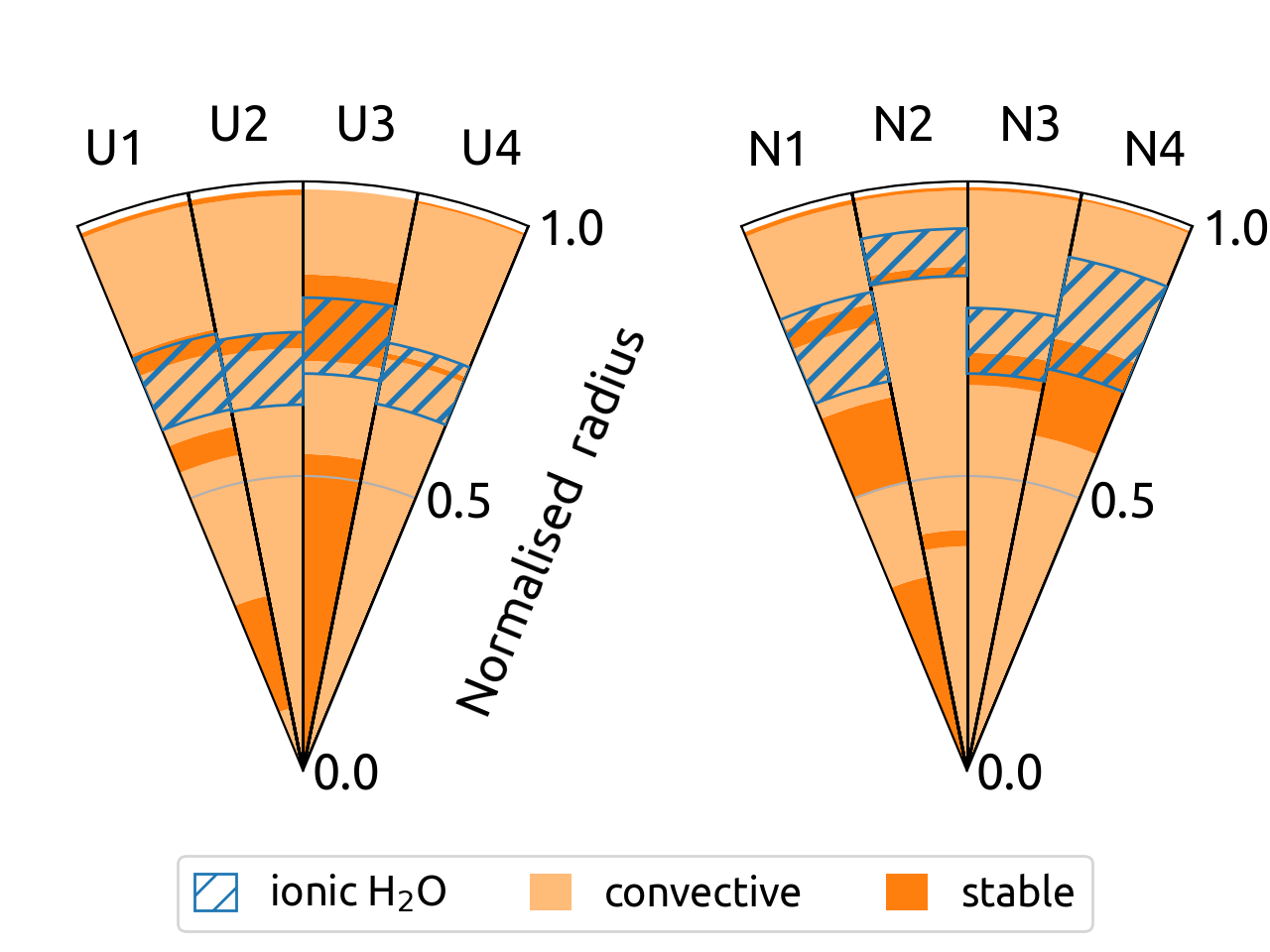}
    \caption{
    Convective and stable regions in our models for Uranus and Neptune.
    Also shown are regions where ionic water is expected. 
    }
    \label{fig:convective_ionic_water}
\end{figure}

%%%%%%%%%%%%%%%%%%%%%%%%%%%%%%%%%%%%%%%%%%%%%%%%%%%%%%%%%%%%%%%%%%%%%%%%%%%%%%%%%%%%%%%%%%

\begin{table}
\caption{
Dynamo regions derived in this study (and compared to Militzer, 2024).
}
\centering
\begin{tabular}{llll}
\hline
\hline
\multicolumn{4}{c}{Uranus} \\
\hline
 & $r_1$ & $r_2$ & $r_1/r_2$ \\
This study & 0.62 -- 0.67 & 0.69 -- 0.74 & 0.86 -- 0.97 \\
\cite{Militzer2024} & 0.49 & 0.73 & 0.67 \\
 & $r_\mu$ & length \\
\cite{Morf2024} & $\sim 0.7-0.9$ & $\lesssim 0.2$ & \\
\hline
\multicolumn{4}{c}{Neptune} \\
\hline
 & $r_1$ & $r_2$ & $r_1/r_2$ \\
This study & 0.67 -- 0.86 & 0.78 -- 0.92 & 0.82 -- 0.93 \\
\cite{Militzer2024} & 0.52 & 0.79 & 0.66 \\
\hline
\hline
\end{tabular}
\tablefoot{
All numbers are given in units of normalised radii.
$r_1$ and $r_2$ denote the lower and upper boundaries of the dynamo regions, respectively.
$r_\mu$ is the mean radius of the dynamo region.
Together with the length of the dynamo region, $r_\mu$ contains the same information as providing $r_1$ and $r_2$.
}
\label{tab:dynamo}
\end{table}

%%%%%%%%%%%%%%%%%%%%%%%%%%%%%%%%%%%%%%%%%%%%%%%%%%%%%%%%%%%%%%%%%%%%%%%%%%%%%%%%%%%%%%%%%%

Figure \ref{fig:convective_ionic_water} shows the extent of the convective and stable regions, as well as the pressure-temperature region where (pure) water is expected to be ionic. 
We find that all the models indeed possess a thin layer of electrically conducting ionic water that is convective. 
We list the upper ($r_2$) and lower ($r_1$) limits of these dynamo regions in Table \ref{tab:dynamo} in units of normalised radii and compare our results to previous work.
For Uranus, we find that our dynamo regions are located deeper in the interior compared to \cite{Morf2024}.
This can be explained by the less steep temperature gradients in this work, leading to a later onset of the ionic phase transition.
Our results for $r_2$ (0.69--0.74 for Uranus and 0.78--0.92 for Neptune) are in agreement with \cite{Militzer2024}.
However, $r_1$ is found to be larger: 
0.62--0.67 for Uranus and 0.67--0.86 for Neptune.
This in turn leads a higher $r_1/r_2$ ratio: 
0.86--0.97 for Uranus and 0.82--0.93 for Neptune, compared to \cite{Militzer2024} and the preferred scenarios in \cite{Stanley2004, Stanley2006}. 
However, our findings for $r_1$ should be interpreted as upper bound values.
This is because the region where the magnetic field is generated can change when considering different composition components. 
Here we only consider pure water and follow the phase diagram of \citet{Redmer2011}.
In principle, other mixtures (such as a mixture of water and rock) could enable dynamo generation \cite[for example][]{Huang2020, Gao2022}. 
In that case our magnetic dynamo regions could extend deeper into the interior than indicated in Table \ref{tab:dynamo}.
This would bring our ratio of $r_1/r_2$ closer to \cite{Stanley2004, Stanley2006, Militzer2024}.
We note that a mixture of rocks with hydrogen or mixtures of other volatiles such as ammonia and methane could also be electronically conducting.
Furthermore, other studies \citep[for example][]{Soderlund2013} found multipolar dynamos while considering different geometries and despite the presence of large, solid, electrically conducting inner cores. 
A better understanding of planetary dynamos as well as material properties at planetary conditions for different elements and mixtures is clearly desirable. 

%%%%%%%%%%%%%%%%%%%%%%%%%%%%%%%%%%%%%%%%%%%%%%%%%%%%%%%%%%%%%%%%%%%%%%%%%%%%%%%%%%%%%%%%%%

\subsection{Immiscibility of hydrogen, helium and water mixtures}
\label{sec:demixing}

%%%%%%%%%%%%%%%%%%%%%%%%%%%%%%%%%%%%%%%%%%%%%%%%%%%%%%%%%%%%%%%%%%%%%%%%%%%%%%%%%%%%%%%%%%

Our models are based on the ideal mixing approximation (Appendix \ref{sec:compositional_algorithm} and Equation \ref{eq:ideal_mixing} contain more details). 
This assumption can lead to differences in density of the order of a few percent \citep[for example][]{Darafeyeu2024}.
In particular, under this assumption immiscibility between different elements is not considered.
\cite{Howard2025} provide miscibility results based on ab initio calculations of the hydrogen-water phase diagram.
However, since our models include a mixture of hydrogen, helium, and water, a correction is needed. 
Using their Equation A.1, we can calculate the pressure-temperature line, below which hydrogen, helium, and water become immiscible.
We employ:
\begin{equation}
    x_{\text{H}_2\text{O}} = \frac{Z_{\text{H}_2\text{O}}\left(1+\frac{\alpha}{2}\right)}{9-Z_{\text{H}_2\text{O}}\left(9-\left(1+\frac{\alpha}{2}\right)\right)}, 
    \label{eq:number_fraction_water}
\end{equation}
when calculating the number fraction of water $x_{\text{H}_2\text{O}}$ from the mass fraction of water $Z_{\text{H}_2\text{O}}$.
The variable $\alpha=0.705/0.275$ denotes the proto-solar mass ratio between hydrogen and helium.
Equation \ref{eq:number_fraction_water} ensures that $x_{\text{H}_2\text{O}} + x_{\text{H}_2} = 1 - x_{\text{He}}$, $m_{\text{H}_2\text{O}}/m_{\text{H}_2} = 9$, and $m_{\text{He}}/m_{\text{H}_2} = 2$.
The inferred demixing lines are presented with dashed lines in Figures \ref{fig:Uranus_densities_TvsP} and \ref{fig:Neptune_densities_TvsP}.
We find that in all models, the $P-T$ lines remain above the demixing curves.
We can therefore conclude that our models are valid, and no demixing is expected. 

%%%%%%%%%%%%%%%%%%%%%%%%%%%%%%%%%%%%%%%%%%%%%%%%%%%%%%%%%%%%%%%%%%%%%%%%%%%%%%%%%%%%%%%%%%

\subsection{Uranus versus Neptune}
\label{sec:compare_uranus_neptune}

%%%%%%%%%%%%%%%%%%%%%%%%%%%%%%%%%%%%%%%%%%%%%%%%%%%%%%%%%%%%%%%%%%%%%%%%%%%%%%%%%%%%%%%%%%

Uranus and Neptune are often thought of as being similar, but it remains unclear how alike these planets truly are \citep[for example][]{Helled2020}.
There are obvious differences between the planets (equatorial radii, masses, rotation periods, gravity fields), which are listed in Table \ref{tab:measured_data}.

%%%%%%%%%%%%%%%%%%%%%%%%%%%%%%%%%%%%%%%%%%%%%%%%%%%%%%%%%%%%%%%%%%%%%%%%%%%%%%%%%%%%%%%%%%

In the outermost convective layers, we tend to find higher H-He mass fractions for Uranus (0.62--0.73) compared to Neptune (0.25--0.49).
This is consistent with the findings of \cite{Nettelmann2013, CanoAmoros2024, Arevalo2025} and could indicate that Uranus indeed possesses more H-He in the outermost regions compared to Neptune.
Furthermore, the outermost edge $r_2$ (Table \ref{tab:dynamo}) of the magnetic dynamo regions tend to be slightly higher up for Neptune (0.78--0.92) compared to Uranus (0.69--0.74). 
This is consistent with the findings of \cite{Militzer2024} and therefore also likely indicative of a dichotomy between Uranus and Neptune. 
Atmospheric probes sent to the atmospheres of both planets could test these predictions. 
Regarding the general bulk compositions (Figure \ref{fig:bulk_compositions}) or the inferred rock-to-water ratios (Table \ref{tab:rock_to_water}), we remain agnostic about inferring concrete differences between Uranus and Neptune.
We can only confirm that a wide range of options are possible for both planets.

%%%%%%%%%%%%%%%%%%%%%%%%%%%%%%%%%%%%%%%%%%%%%%%%%%%%%%%%%%%%%%%%%%%%%%%%%%%%%%%%%%%%%%%%%%

We further note that the two planets have different measured heat fluxes \citep[][]{Pearl1991} which is often explained with different structures such as gradients or boundary layers \citep[for example][]{Vazan2020, Scheibe2019, Scheibe2021} or giant impacts \citep[for example][]{Reinhardt2020}.
Our models do not include energy balance or flux calculations. 
If differences in structures are indeed the reason for the different heat fluxes, that would favour models U1 and U3 for Uranus and models N2 and N3 for Neptune.
Models U1 and U3 contain extensive composition gradients where heat transport can not be facilitated efficiently via convection.
As a consequence, comparatively little heat would be able to escape Uranus, while the large convection zones in models N2 and N3 would enable a more efficient heat transport and therefore a larger heat flux in comparison to Uranus.
Interestingly, models U1 and U3 are rock-poor, while models N2 and N3 are rock-rich.
While this could be indicative of a bulk composition dichotomy between Uranus and Neptune, we refrain from making any definite statements.
For a robust conclusion, both more models and heat flux calculations (and measurements) would be required.
We hence remain agnostic about this issue and hope to investigate this topic in detail in future work.

%%%%%%%%%%%%%%%%%%%%%%%%%%%%%%%%%%%%%%%%%%%%%%%%%%%%%%%%%%%%%%%%%%%%%%%%%%%%%%%%%%%%%%%%%%
%%%%%%%%%%%%%%%%%%%%%%%%%%%%%%%%%%%%%%%%%%%%%%%%%%%%%%%%%%%%%%%%%%%%%%%%%%%%%%%%%%%%%%%%%%

\section{Discussion}
\label{sec:discussion}

%%%%%%%%%%%%%%%%%%%%%%%%%%%%%%%%%%%%%%%%%%%%%%%%%%%%%%%%%%%%%%%%%%%%%%%%%%%%%%%%%%%%%%%%%%

This work introduces a novel method for modelling the internal structures of planetary interiors. 
By uniting the flexibility of empirical models with the physical validity of traditional approaches, we produce agnostic and self-consistent interior solutions.
Our iterative algorithm allows for the convergence of randomly initialised density profiles.
They converge towards solutions that simultaneously satisfy hydrostatic equilibrium, fit the gravity data, and respect constraints imposed by the employed EoS. 
We note that interior models can be further constrained by seismology  \cite{Mankovich2025}.
We provide realistic Brunt-Väisälä frequencies necessary for interpreting oscillation modes in planetary interiors in Figures \ref{fig:physical_solutions_temperatures_entropies_N2_gen_4_U} and \ref{fig:physical_solutions_temperatures_entropies_N2_gen_4_N}.
Our models are well suited for comparison with asteroseismic observations by future missions. 
While this study represents a major step forward in planetary modelling, some limitations remain and future research is clearly desirable. 

%%%%%%%%%%%%%%%%%%%%%%%%%%%%%%%%%%%%%%%%%%%%%%%%%%%%%%%%%%%%%%%%%%%%%%%%%%%%%%%%%%%%%%%%%%

First, the EoS of materials at planetary conditions themselves may contain errors that can affect the results \citep[][]{More1988, Chabrier2021, Haldemann2020, Cozza2025}. 
For example, the calculations in \cite{Haldemann2020} contain an error that affects the entropy, but not its derivatives \citep{Aguichine2025}.
\cite{Arevalo2025} find that correcting for this error produces qualitatively similar results.
All EoSs of pure materials possess uncertainties and it is hence clear that the choice of the EoS affects the results.
In Appendix \ref{sec:EoS_comparison}, we compare two different EoS for SiO$_2$ to quantify how different their results are as an additional example.
We find that the differences in density can be of the order of a few percent.
Furthermore, the assumption of ideal mixing (Equation \ref{eq:ideal_mixing}) can yet again lead to differences in density of the order of a few percent \citep[for example][]{Darafeyeu2024} and therefore affect the inferred composition. 
So even if the EoSs of the pure materials would serve as perfect constraints, a complexity arises from the fact that the mixtures are treated in an over-simplified manner.
Finally, we only considered immiscibility for the hydrogen-helium-water mixture, but did not include other immiscibilities (like a mixture of water and rock). 
Overall, we can therefore expect that the uncertainties associated with all of the aforementioned issues would also affect the inferred bulk compositions.

%%%%%%%%%%%%%%%%%%%%%%%%%%%%%%%%%%%%%%%%%%%%%%%%%%%%%%%%%%%%%%%%%%%%%%%%%%%%%%%%%%%%%%%%%%

Second, for the elements included in Uranus and Neptune, we only considered hydrogen, helium, water, rocks, and iron. 
Other elements such as methane or ammonia are also expected to exist in the planetary interiors and could also influence the results \citep[for example][]{Malamud2024,Arevalo2025}.
For example, methane and ammonia, being less dense than water and rock, can contribute to the reduced densities required by gravitational harmonic constraints.
Consequently, our models may over-predict the hydrogen and helium abundance.
Including methane and ammonia could also elevate the temperature profiles due to these lower densities, assuming a constant specific entropy.
In this work we do not include methane and ammonia since the exact abundance ratios between different elements remain uncertain.
In other words, one would either have to rely on an educated guess \citep[][]{Arevalo2025}, or add even more free parameters to the problem.

%%%%%%%%%%%%%%%%%%%%%%%%%%%%%%%%%%%%%%%%%%%%%%%%%%%%%%%%%%%%%%%%%%%%%%%%%%%%%%%%%%%%%%%%%%

Third, our models rely on the pressure-density relation provided in \cite{Hueso2020} for pressures below 100 bars.
We rely on an atmospheric model because our approach is not well suited to handle the complex chemical processes occurring in planetary atmospheres such as condensation.
Atmospheric models are generally more detailed and therefore more accurate in representing the conditions in the planetary atmosphere.
Therefore, we rather not assume an atmospheric composition that could bias the results of the interior model.
However, we do ensure that our models have density-pressure relations that are compatible with these more accurate atmospheric models \citep[][for details]{Morf2024}.
Although the used atmosphere model by \cite{Hueso2020} should be more accurate than our approach below 100 bars, it also includes uncertainties.
To account for this, we allowed the initial temperature $T^0_\text{EoS}$ to deviate from the atmosphere model.
Nonetheless, a different atmosphere model could lead to different results.

%%%%%%%%%%%%%%%%%%%%%%%%%%%%%%%%%%%%%%%%%%%%%%%%%%%%%%%%%%%%%%%%%%%%%%%%%%%%%%%%%%%%%%%%%%
    
Fourth, we did not consider dynamical wind corrections to the measured gravitational moments \citep{Kaspi2013, Soyuer2023}.
This can affect the density profile solution space and therefore our findings.
However, \cite{Neuenschwander2024} investigate this effect for Uranus and suggest that wind corrections should not have a large influence on the inferred temperature and composition profiles.
Since Uranus' gravity data are significantly more constrained than Neptune's, we expect that the same holds true for Neptune.

%%%%%%%%%%%%%%%%%%%%%%%%%%%%%%%%%%%%%%%%%%%%%%%%%%%%%%%%%%%%%%%%%%%%%%%%%%%%%%%%%%%%%%%%%%

Finally, the rotation periods of Uranus and Neptune might deviate from the rotation periods listed in Table \ref{tab:measured_data} \citep[][]{Helled2010}. 
Changing the rotation periods has a large impact on the density profile solutions space and the inferred composition \citep[for example][]{Nettelmann2013, Neuenschwander2024}.  
For this work, we rely on purely observational constraints \citep{Desch1986, Karkoschka2011} as there is no broader consensus or further validation of the alternative estimates by \cite{Helled2010}.
We acknowledge the importance of this issue and plan to explore it in future work.

%%%%%%%%%%%%%%%%%%%%%%%%%%%%%%%%%%%%%%%%%%%%%%%%%%%%%%%%%%%%%%%%%%%%%%%%%%%%%%%%%%%%%%%%%%
%%%%%%%%%%%%%%%%%%%%%%%%%%%%%%%%%%%%%%%%%%%%%%%%%%%%%%%%%%%%%%%%%%%%%%%%%%%%%%%%%%%%%%%%%%

\section{Conclusions}
\label{sec:conclusions}

%%%%%%%%%%%%%%%%%%%%%%%%%%%%%%%%%%%%%%%%%%%%%%%%%%%%%%%%%%%%%%%%%%%%%%%%%%%%%%%%%%%%%%%%%%

The key conclusions of our study can be summarised as follows: 

%%%%%%%%%%%%%%%%%%%%%%%%%%%%%%%%%%%%%%%%%%%%%%%%%%%%%%%%%%%%%%%%%%%%%%%%%%%%%%%%%%%%%%%%%%

\begin{itemize}

%%%%%%%%%%%%%%%%%%%%%%%%%%%%%%%%%%%%%%%%%%%%%%%%%%%%%%%%%%%%%%%%%%%%%%%%%%%%%%%%%%%%%%%%%%
    
    \item Bridging empirical and physical interior modelling: \vspace{0.1cm} \\
    By using both random density profiles and a random compositional algorithm, we minimise bias for our planetary interior models.
    At the same time, we ensure a level of self-consistency typically reserved for physical interior models by using a novel iterative procedure.
    Our results thus cover a wide range of options for interior density, temperature, and composition profiles that are physical and robust.
    
%%%%%%%%%%%%%%%%%%%%%%%%%%%%%%%%%%%%%%%%%%%%%%%%%%%%%%%%%%%%%%%%%%%%%%%%%%%%%%%%%%%%%%%%%%
    
    \item \vspace{0.1cm} Uranus and Neptune as 'rock giants': \vspace{0.1cm} \\
    We find that different bulk compositions are consistent with current data for both Uranus and Neptune.
    They encompass rock-dominated and water-rich interiors, with rock-to-water ratios varying by more than an order of magnitude (0.04--3.92 for Uranus, 0.20--1.78 for Neptune; Table \ref{tab:rock_to_water}). 
    We hence confirm that the label 'ice giants' for Uranus and Neptune may be more of a historical artifact rather than a robust physical classification. 

%%%%%%%%%%%%%%%%%%%%%%%%%%%%%%%%%%%%%%%%%%%%%%%%%%%%%%%%%%%%%%%%%%%%%%%%%%%%%%%%%%%%%%%%%%
    
    \item \vspace{0.1cm} Differences between Uranus and Neptune: \vspace{0.1cm} \\
    In the outermost convective zones, we find a higher local H-He mass fraction for Uranus (0.62--0.73) than for Neptune (0.25--0.49).
    Furthermore, it seems that Uranus' magnetic field is generated deeper in the planetary interior compared to Neptune.
    We find upper limits of 0.69--0.74 (Uranus) vs. 0.78--0.92 (Neptune) for the outer edges of the dynamo regions in units of normalised radii (Table \ref{tab:dynamo}).
    
%%%%%%%%%%%%%%%%%%%%%%%%%%%%%%%%%%%%%%%%%%%%%%%%%%%%%%%%%%%%%%%%%%%%%%%%%%%%%%%%%%%%%%%%%%

    \item \vspace{0.1cm} Dynamo and immiscibility constraints matched: \vspace{0.1cm} \\
    All our models contain convective layers where (pure) water exists in its ionic phase. 
    Such regions could serve as the dynamo source of Uranus' and Neptune’s multipolar magnetic fields. 
    Furthermore, none of the modelled temperature–pressure profiles cross the hydrogen–water immiscibility boundaries derived from recent ab-initio calculations.

%%%%%%%%%%%%%%%%%%%%%%%%%%%%%%%%%%%%%%%%%%%%%%%%%%%%%%%%%%%%%%%%%%%%%%%%%%%%%%%%%%%%%%%%%%

\end{itemize}

%%%%%%%%%%%%%%%%%%%%%%%%%%%%%%%%%%%%%%%%%%%%%%%%%%%%%%%%%%%%%%%%%%%%%%%%%%%%%%%%%%%%%%%%%%

With the potential for future dedicated missions to Uranus and Neptune, our method also provides a flexible and unbiased tool for interpreting forthcoming data.
Ultimately, the interiors of Uranus and Neptune remain enigmatic, not because they are beyond reach, but because the data required to resolve their secrets are still out of grasp. 
Until then, only a plurality of models, not a singular one, can capture the full extent of possibilities for their hidden depths.

%%%%%%%%%%%%%%%%%%%%%%%%%%%%%%%%%%%%%%%%%%%%%%%%%%%%%%%%%%%%%%%%%%%%%%%%%%%%%%%%%%%%%%%%%%
%%%%%%%%%%%%%%%%%%%%%%%%%%%%%%%%%%%%%%%%%%%%%%%%%%%%%%%%%%%%%%%%%%%%%%%%%%%%%%%%%%%%%%%%%%

\begin{acknowledgements}

We thank the anonymous referee for his valuable suggestions that helped strengthen this work. 
We also would like to thank Simon Müller and Christian Reinhardt for providing us with numerical implementations of the employed EoS. 
This work was supported by the Swiss National Science Foundation (SNSF) through a grant provided as a part of project number 215634: \url{https://data.snf.ch/grants/grant/215634}.
      
\end{acknowledgements}

%%%%%%%%%%%%%%%%%%%%%%%%%%%%%%%%%%%%%%%%%%%%%%%%%%%%%%%%%%%%%%%%%%%%%%%%%%%%%%%%%%%%%%%%%%
%%%%%%%%%%%%%%%%%%%%%%%%%%%%%%%%%%%%%%%%%%%%%%%%%%%%%%%%%%%%%%%%%%%%%%%%%%%%%%%%%%%%%%%%%%

\bibliographystyle{aa}
\bibliography{literature.bib}

%%%%%%%%%%%%%%%%%%%%%%%%%%%%%%%%%%%%%%%%%%%%%%%%%%%%%%%%%%%%%%%%%%%%%%%%%%%%%%%%%%%%%%%%%%%%%%%%%%%%%%%%%%%%%%%%%%%%%%%%%%%%%%%%%%%%%%%%%%%%%%%%%%%%%%%%%%%%%%%%%%%%%%%%%%%%%%%%%%%%

\begin{appendix}

%%%%%%%%%%%%%%%%%%%%%%%%%%%%%%%%%%%%%%%%%%%%%%%%%%%%%%%%%%%%%%%%%%%%%%%%%%%%%%%%%%%%%%%%%%%%%%%%%%%%%%%%%%%%%%%%%%%%%%%%%%%%%%%%%%%%%%%%%%%%%%%%%%%%%%%%%%%%%%%%%%%%%%%%%%%%%%%%%%%%

\section{Additional interior data}
\label{sec:additional_data}

%%%%%%%%%%%%%%%%%%%%%%%%%%%%%%%%%%%%%%%%%%%%%%%%%%%%%%%%%%%%%%%%%%%%%%%%%%%%%%%%%%%%%%%%%%

\begin{figure}
    \centering
    \includegraphics[width=\hsize]{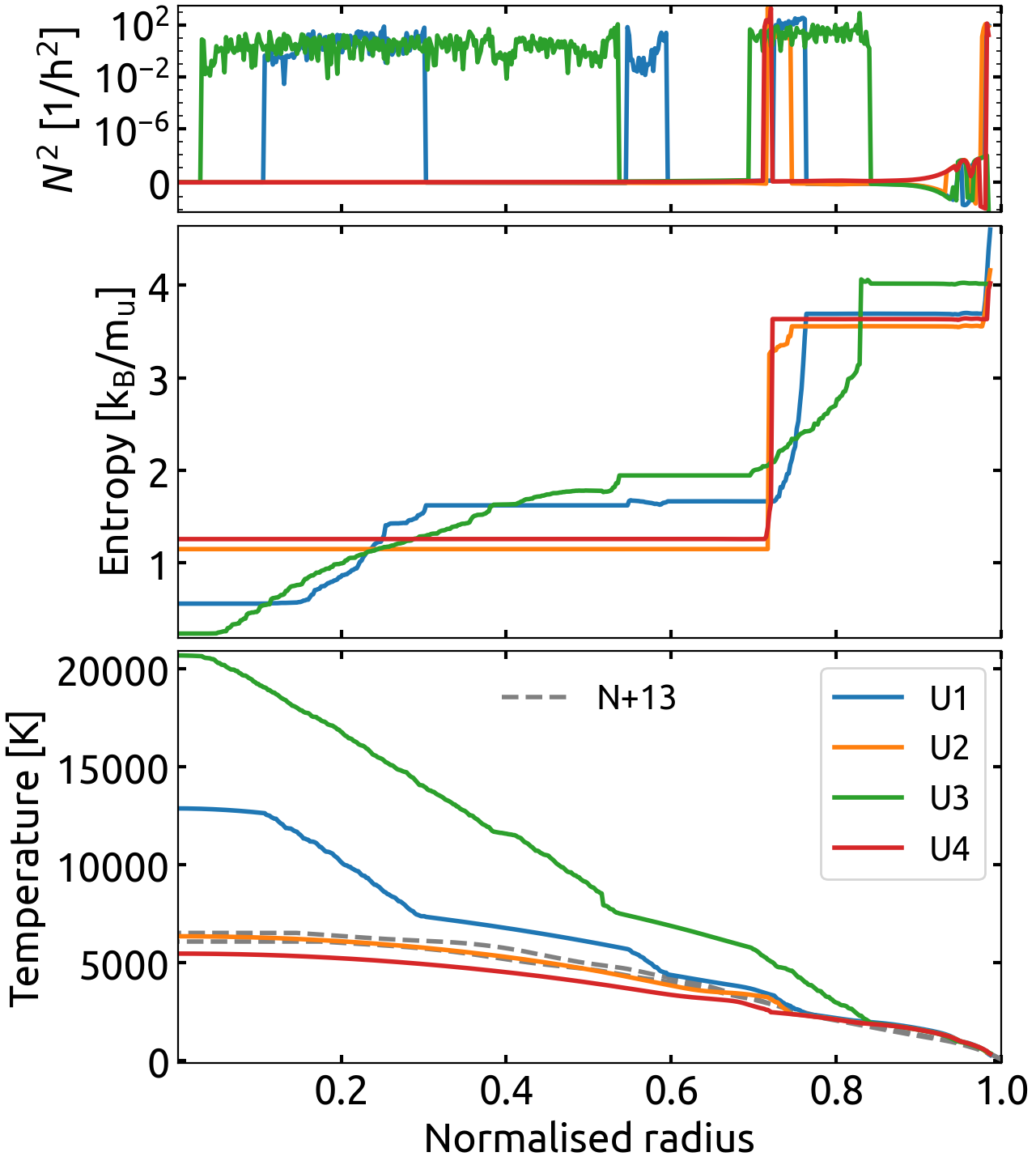}
    \caption{
    Brunt-Väisälä frequencies (top), entropies (middle), and temperature (bottom) profiles of the four Uranus models.
    For comparison, we also show purely adiabatic temperature profiles from \cite{Nettelmann2013} (dashed grey).
    }
\label{fig:physical_solutions_temperatures_entropies_N2_gen_4_U}
\end{figure}

%%%%%%%%%%%%%%%%%%%%%%%%%%%%%%%%%%%%%%%%%%%%%%%%%%%%%%%%%%%%%%%%%%%%%%%%%%%%%%%%%%%%%%%%%%

\begin{figure}
    \centering
    \includegraphics[width=\hsize]{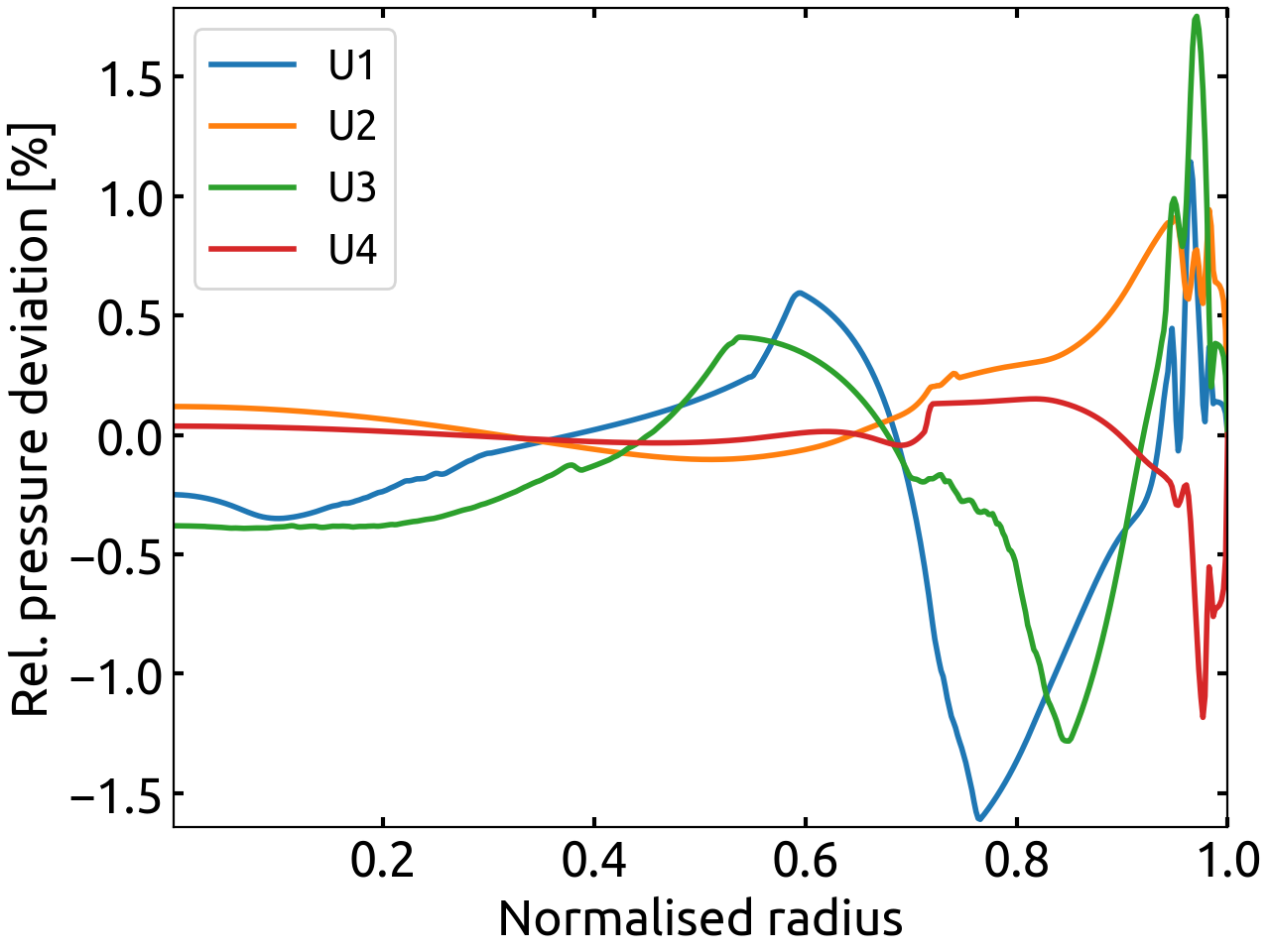}
    \caption{
    Relative pressure deviations (Equation \ref{eq:pressure_conistency}) of the four Uranus models.
    }
    \label{fig:physical_solutions_pressure_deltas_gen_4_U}
\end{figure}

%%%%%%%%%%%%%%%%%%%%%%%%%%%%%%%%%%%%%%%%%%%%%%%%%%%%%%%%%%%%%%%%%%%%%%%%%%%%%%%%%%%%%%%%%%

\begin{figure}
    \centering
    \includegraphics[width=\hsize]{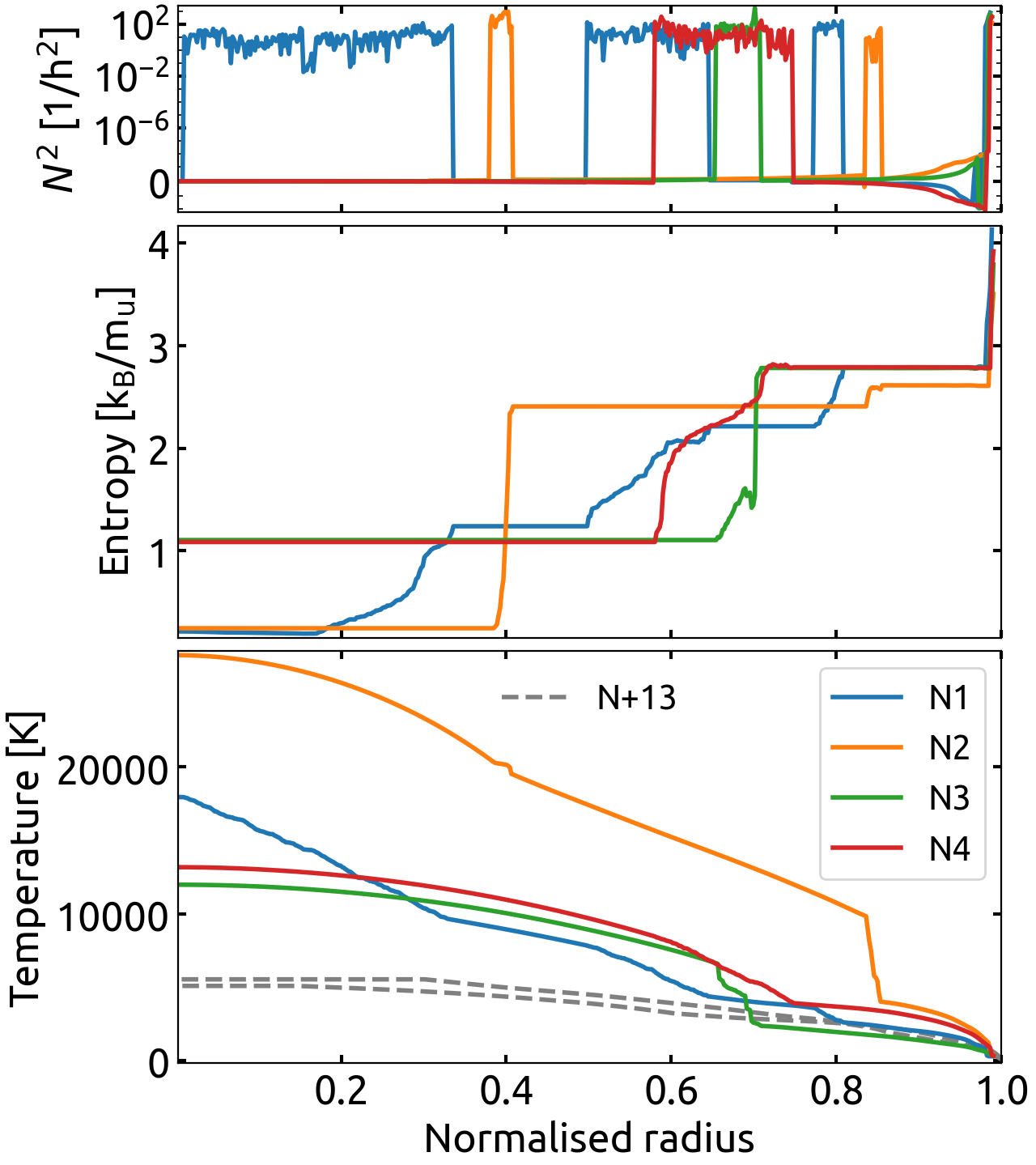}
    \caption{
    Same as Figure \ref{fig:physical_solutions_temperatures_entropies_N2_gen_4_U}, but for Neptune.
    }
\label{fig:physical_solutions_temperatures_entropies_N2_gen_4_N}
\end{figure}

%%%%%%%%%%%%%%%%%%%%%%%%%%%%%%%%%%%%%%%%%%%%%%%%%%%%%%%%%%%%%%%%%%%%%%%%%%%%%%%%%%%%%%%%%%

\begin{figure}
    \centering
    \includegraphics[width=\hsize]{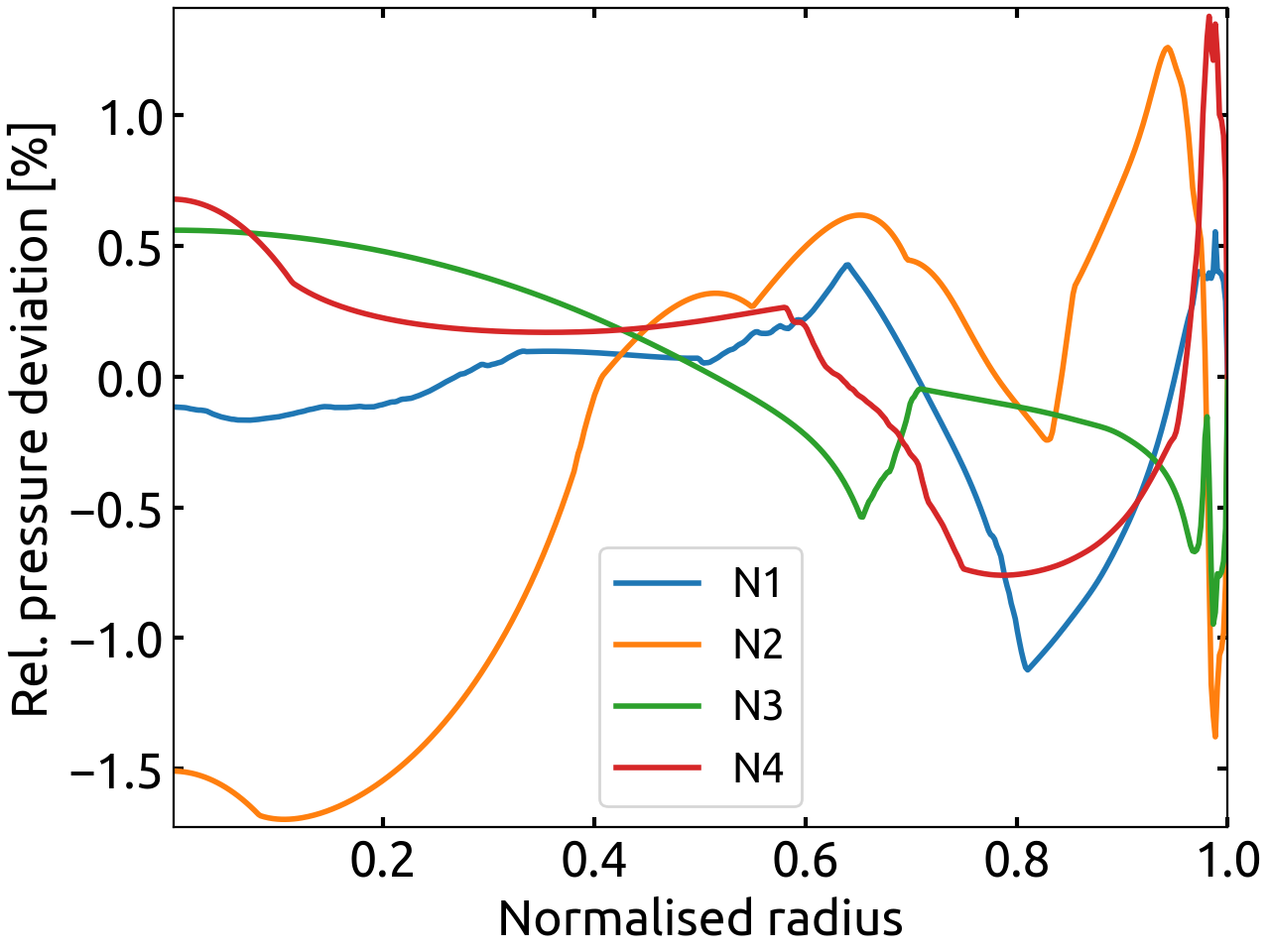}
    \caption{
    Same as Figure \ref{fig:physical_solutions_pressure_deltas_gen_4_U}, but for Neptune.
    }
    \label{fig:physical_solutions_pressure_deltas_gen_4_N}
\end{figure}

%%%%%%%%%%%%%%%%%%%%%%%%%%%%%%%%%%%%%%%%%%%%%%%%%%%%%%%%%%%%%%%%%%%%%%%%%%%%%%%%%%%%%%%%%%

We show interior temperature, entropy, and frequency profiles for the models of Uranus and Neptune in Figures \ref{fig:physical_solutions_temperatures_entropies_N2_gen_4_U} and \ref{fig:physical_solutions_temperatures_entropies_N2_gen_4_N}.
We find that, as expected, the Brunt-Väisälä frequency $N$ is nearly zero ($|N^2|<10^{-8}\text{h}^{-2}$) in adiabatic regions. 
This is orders of magnitudes smaller than analogous results in \cite{Morf2024}.
The frequency $N^2$ should be smaller than zero throughout all adiabatic regions subjected to convection.
This is violated in the outermost parts (normalised radii above 0.9) in Figures \ref{fig:physical_solutions_temperatures_entropies_N2_gen_4_U} and \ref{fig:physical_solutions_temperatures_entropies_N2_gen_4_N}.
However, these are caused by numerical floating point precision limitations with near-zero numbers, and they do not contain any physical meaning.
The wiggles visible in Ledoux-stable regions ($N^2>10^{-2}\text{h}^{-2}$) originate from the random nature of the algorithm.

%%%%%%%%%%%%%%%%%%%%%%%%%%%%%%%%%%%%%%%%%%%%%%%%%%%%%%%%%%%%%%%%%%%%%%%%%%%%%%%%%%%%%%%%%%

We show the relative pressure deviation criterion introduced in Equation \ref{eq:pressure_conistency} in Figure \ref{fig:physical_solutions_pressure_deltas_gen_4_U} for Uranus and Figure \ref{fig:physical_solutions_pressure_deltas_gen_4_N} for Neptune.
We find that all the models are consistent with the $\epsilon=0.02$ limit introduced in Section \ref{sec:methods}.
In this work, the number $k$ in Equation \ref{eq:pressure_conistency} ranges from 3 to 8. 

%%%%%%%%%%%%%%%%%%%%%%%%%%%%%%%%%%%%%%%%%%%%%%%%%%%%%%%%%%%%%%%%%%%%%%%%%%%%%%%%%%%%%%%%%%

\renewcommand{\thefigure}{B.\arabic{figure}}
\setcounter{figure}{0}

\begin{figure*}[!htbp]
    \centering
    \includegraphics[width=\hsize]{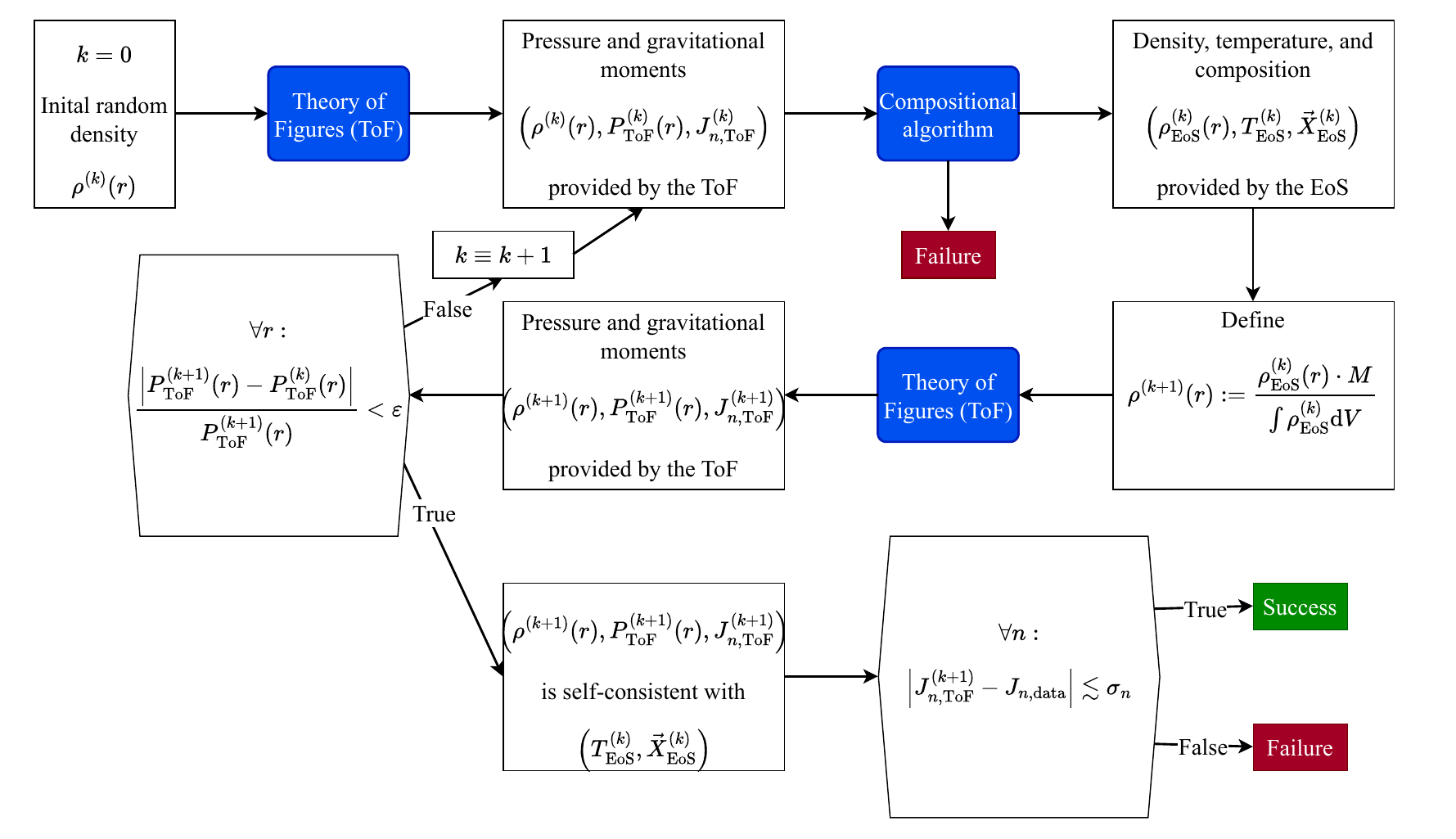}
    \caption{
    Global algorithm with more details than Figure \ref{fig:high_level_overview_algorithm}.
    As input, it requires a density profile $\rho^{(0)}(r)$ of a planetary interior.
    In this work, $\rho^{(0)}(r)$ is generated in a random manner.
    The global algorithm then repeatedly calls the Theory of Figures (ToF) and a compositional algorithm to infer the  pressure, temperature, and composition profile of the planet.
    The algorithm is applied until everything is self-consistent up to a certain tolerance.
    The algorithm is then deemed successful if the gravitational moments calculated by the ToF are compatible with the provided data.
    }
    \label{fig:global_algorithm}
\end{figure*}

%%%%%%%%%%%%%%%%%%%%%%%%%%%%%%%%%%%%%%%%%%%%%%%%%%%%%%%%%%%%%%%%%%%%%%%%%%%%%%%%%%%%%%%%%%

\begin{figure*}
    \centering
    \includegraphics[width=\hsize]{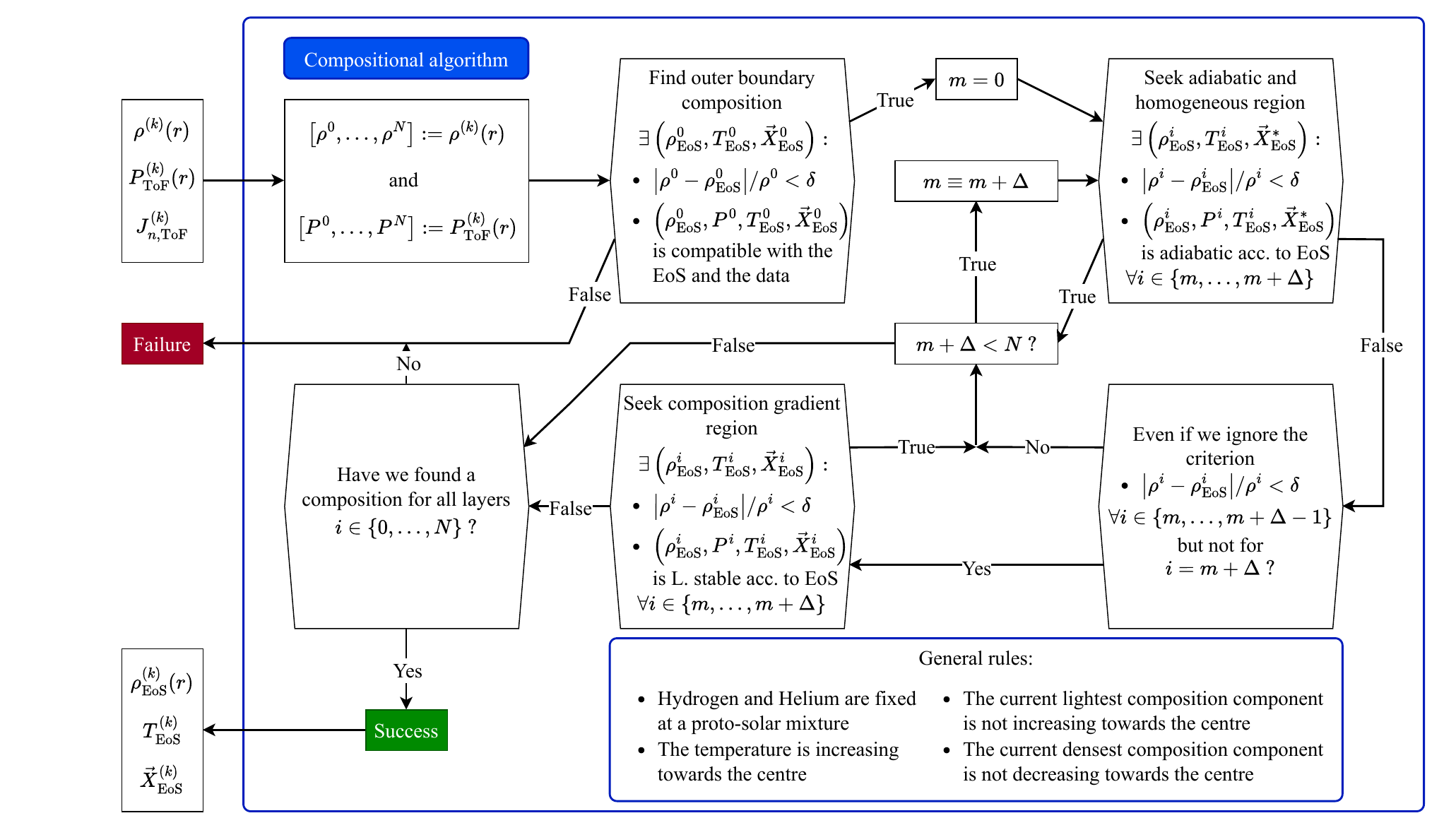}
    \caption{
    Compositional algorithm. 
    As input, it requires density values $[\rho^0, \dots, \rho^N]$ and pressure values $[P^0, \dots, P^N]$ for the planetary interior. 
    If successful, the algorithm infers a temperature and composition profile. 
    The densities implied by the Equations of State (EoS) can deviate from the input densities $[\rho^0, \dots, \rho^N]$.
    }
    \label{fig:compositional_algorithm}
\end{figure*}

%%%%%%%%%%%%%%%%%%%%%%%%%%%%%%%%%%%%%%%%%%%%%%%%%%%%%%%%%%%%%%%%%%%%%%%%%%%%%%%%%%%%%%%%%%
%%%%%%%%%%%%%%%%%%%%%%%%%%%%%%%%%%%%%%%%%%%%%%%%%%%%%%%%%%%%%%%%%%%%%%%%%%%%%%%%%%%%%%%%%%

\section{Compositional algorithm}
\label{sec:compositional_algorithm}

%%%%%%%%%%%%%%%%%%%%%%%%%%%%%%%%%%%%%%%%%%%%%%%%%%%%%%%%%%%%%%%%%%%%%%%%%%%%%%%%%%%%%%%%%%

We now discuss the compositional algorithm in detail, a part of the global algorithm depicted in Figure \ref{fig:global_algorithm}. 
The compositional algorithm is depicted in Figure \ref{fig:compositional_algorithm} and represents an improved version of the work presented in \cite{Morf2024}.
As input, it requires both density values $[\rho^0, \dots, \rho^N]$ and pressure values $[P^0, \dots, P^N]$ within the planet.
The index $0$ denotes the outermost layer, where $N$ corresponds to the planetary centre.
In this work, we defined the first layer that has a pressure above 100 bars as the outermost layer.
For layers below a pressure of 100 bar, we did not infer a composition and just fixed the profile to follow the density-pressure atmosphere models proposed by \cite{Hueso2020}.
This is motivated by more complex chemical processes in the atmosphere \citep[such as in][]{Hueso2020}, which our approach is not well suited for. 

%%%%%%%%%%%%%%%%%%%%%%%%%%%%%%%%%%%%%%%%%%%%%%%%%%%%%%%%%%%%%%%%%%%%%%%%%%%%%%%%%%%%%%%%%%

We considered hydrogen ($X$) and helium ($Y$), water ($Z_1$), rocks ($Z_2$), and iron ($Z_3$) for a composition mass fraction $\vec{X}_\text{EoS} = (X,Y,Z_1,Z_2,Z_3)$.
To that end, we relied on the EoS provided by \cite{More1988, Chabrier2021, Haldemann2020}.
The abundance of hydrogen and helium relative to each other was always fixed at a proto-solar mass ratio of $0.705/0.275$.
We assumed the ideal mixing equation
\begin{equation}
    \frac{1}{\rho} = \frac{X}{\rho_X} + \frac{Y}{\rho_Y} + \frac{Z_1}{\rho_{Z_1}} + \frac{Z_2}{\rho_{Z_2}} + \frac{Z_3}{\rho_{Z_3}}, 
    \label{eq:ideal_mixing}
\end{equation}
to combine multiple different materials and their respective EoS.

%%%%%%%%%%%%%%%%%%%%%%%%%%%%%%%%%%%%%%%%%%%%%%%%%%%%%%%%%%%%%%%%%%%%%%%%%%%%%%%%%%%%%%%%%%

In the beginning, the compositional algorithm chooses a random temperature $T_\text{EoS}^0$.
For this work, we randomly chose $T_\text{EoS}^0 \in [250, 450]$ K for the first layer that has a pressure above 100 bars.
This temperature range is based on the results of \cite{Hueso2020}.
Given $T_\text{EoS}^0$, the compositional algorithm tries to find a composition $\vec{X}_\text{EoS}^0$ that results in a density $\rho_\text{EoS}^0$ that is close enough to the provided density $\rho^0$, meaning
\begin{equation}
    \frac{\left|\rho^0-\rho_\text{EoS}^0\right|}{\rho^0} < \delta.
    \label{eq:rho_tolerance}
\end{equation}
Here, we used $\delta = 0.01$.
If multiple such $\rho_\text{EoS}^0$ exist, the algorithm chooses one of them randomly.

%%%%%%%%%%%%%%%%%%%%%%%%%%%%%%%%%%%%%%%%%%%%%%%%%%%%%%%%%%%%%%%%%%%%%%%%%%%%%%%%%%%%%%%%%%

Moving towards the planetary interior, the compositional algorithm prioritises finding regions of constant composition.
These regions are adiabatic due to large scale convection mixing the materials homogeneously.
To find them, the compositional algorithm checks whether it can keep the composition $\vec{X}_\text{EoS}^*$ from the previous layer and follow an adiabatic gradient for the next $\Delta$ layers.
The check is successful if the densities inferred by the EoS obey Equation \ref{eq:rho_tolerance} for all $\Delta$ layers.
We only accepted such regions if $\Delta > N/8 = 64$.
This limit was introduced to prevent small convection cells.
While small convection cells are possible, sustaining them over long enough time scales is tricky and highly dependent on dynamical processes such as oscillatory double-diffusive convection \citep{Leconte2012, Tulekeyev2024}.
We do not investigate such dynamical processes better treated by evolution models \citep[such as][]{Arevalo2025, Eberlein2025}, which motivated the aforementioned limit.

%%%%%%%%%%%%%%%%%%%%%%%%%%%%%%%%%%%%%%%%%%%%%%%%%%%%%%%%%%%%%%%%%%%%%%%%%%%%%%%%%%%%%%%%%%

As a second priority, the compositional algorithm checks if it can find a homogeneous and adiabatic region by ignoring Equation \ref{eq:rho_tolerance} for $\Delta-1$ layers.
Only satisfying Equation \ref{eq:rho_tolerance} again for the $\Delta$-th layer violates self-consistency to a high degree.
However, due to the iterative nature of the global algorithm, the necessity for such violations should become less and less frequent.

%%%%%%%%%%%%%%%%%%%%%%%%%%%%%%%%%%%%%%%%%%%%%%%%%%%%%%%%%%%%%%%%%%%%%%%%%%%%%%%%%%%%%%%%%%

Finally, as a third priority, the algorithm allows the composition to change:
In addition to the inferred $\rho_\text{EoS}$ satisfying Equation \ref{eq:rho_tolerance}, one now needs to check for stability against convection.
This is done by calculating the Brunt-Väisälä frequency $N$.
For an arbitrary density $\rho$, pressure $P$, and temperature $T$, $N$ is given by \citep[for example][]{Unno1989, Kippenhahn2013}
\begin{equation}
    N^2=\frac{g^2 \rho}{P} \frac{\chi_T}{\chi_\rho}\left(\nabla_\text{ad}-\nabla_T+B\right), 
\end{equation}
where $g$ is the gravity acceleration.
The thermodynamic quantities $\chi_\rho$ and $\chi_T$ are defined as
\begin{equation}
\chi_\rho=\left(\frac{\partial \ln P}{\partial \ln \rho}\right)_{T, \mu}, \quad \chi_T=\left(\frac{\partial \ln P}{\partial \ln T}\right)_{\rho, \mu},
\end{equation}
where the molecular weight $\mu$ parametrises the composition. 
In particular, $N$ depends on the difference between the adiabatic gradient $\nabla_\text{ad}$ and temperature gradient $\nabla_T$ 
\begin{equation}
\nabla_\text{ad}:=\left(\frac{\partial \ln T}{\partial \ln P}\right)_S, \quad \nabla_T:=\frac{d \ln T}{d \ln P},
\end{equation}
as well as on the mixing term $B$
\begin{equation}
B=\frac{\chi_\rho}{\chi_T} \frac{\ln \rho\left(P, T, \mu+\frac{d \mu}{d \ln \rho} \delta \ln \rho\right)-\ln \rho(P, T, \mu)}{\delta \ln \rho} \frac{d \ln \rho}{d \ln P},
\end{equation}
taken in the limit $\delta \ln \rho \rightarrow 0$. 
The mixing term $B$ can be approximated numerically \citep[Appendix B of][]{Morf2024}.
The Brunt-Väisälä frequency $N$ denotes the oscillation frequency of a parcel of material that is slightly displaced compared to its surroundings while not exchanging any heat with it.
For stability, one requires $N^2 \geq 0$.
The criterion $N^2 \geq 0$ is called the Ledoux criterion.
If multiple tuples $\left(\rho^i_\text{EoS}, T^i_\text{EoS},\vec{X}^i_\text{EoS}\right)$ satisfy both Equation \ref{eq:rho_tolerance} and $N^2 \geq 0$, the algorithm chooses one of them randomly.

%%%%%%%%%%%%%%%%%%%%%%%%%%%%%%%%%%%%%%%%%%%%%%%%%%%%%%%%%%%%%%%%%%%%%%%%%%%%%%%%%%%%%%%%%%

The compositional algorithm succeeds by repeatedly applying the above steps in accordance with their priorities for the whole planet.
The compositional algorithm fails if at some point none of the above steps are possible.
If successful, it provides a temperature and composition profile for the planet.
We stress again that the densities implied by the EoS can deviate from the input densities, sometimes even more than allowed by Equation \ref{eq:rho_tolerance}.
The inferred temperature and composition profiles are hence only physical, if the entire model (Equation \ref{eq:complete_tuple}) satisfies the relative pressure deviation criterion stated in Equation \ref{eq:pressure_conistency}.

%%%%%%%%%%%%%%%%%%%%%%%%%%%%%%%%%%%%%%%%%%%%%%%%%%%%%%%%%%%%%%%%%%%%%%%%%%%%%%%%%%%%%%%%%%
%%%%%%%%%%%%%%%%%%%%%%%%%%%%%%%%%%%%%%%%%%%%%%%%%%%%%%%%%%%%%%%%%%%%%%%%%%%%%%%%%%%%%%%%%%

\section{Algorithm convergence}
\label{sec:algorithm_convergence}

%%%%%%%%%%%%%%%%%%%%%%%%%%%%%%%%%%%%%%%%%%%%%%%%%%%%%%%%%%%%%%%%%%%%%%%%%%%%%%%%%%%%%%%%%%

\renewcommand{\thefigure}{C.\arabic{figure}}

\begin{figure}
    \centering
    \includegraphics[width=\hsize]{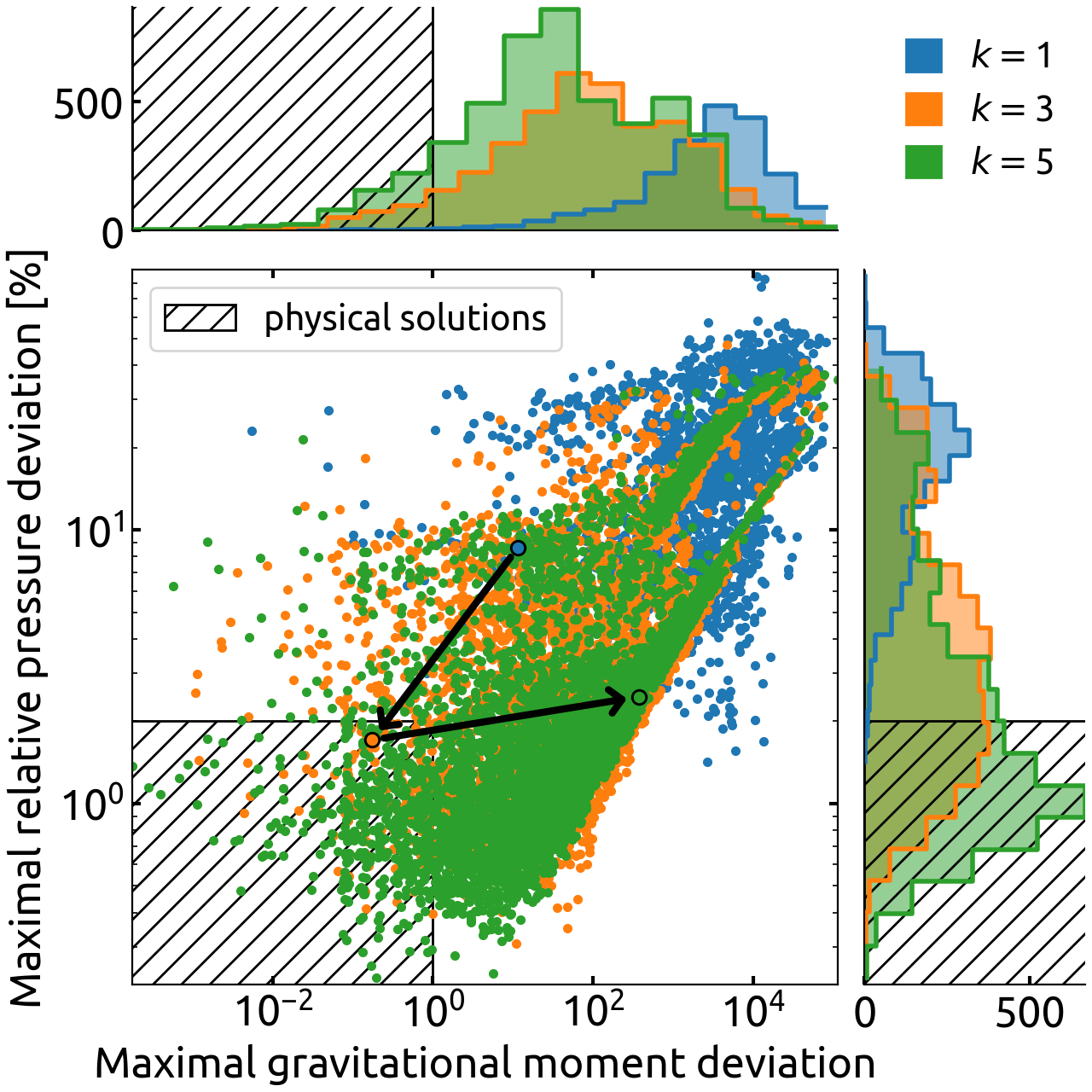}
    \caption{
    Maximal gravitational moment deviation versus maximal relative pressure deviation for different generations $k\in\{1,3,5\}$ and in total over ten thousand Neptune models.
    Histograms of the depicted points are included for both axes.
    Points falling within the hatched area represent physical models.
    The arrows highlight one single path between generations.
    }
\label{fig:convergence_overview_N}
\end{figure}

%%%%%%%%%%%%%%%%%%%%%%%%%%%%%%%%%%%%%%%%%%%%%%%%%%%%%%%%%%%%%%%%%%%%%%%%%%%%%%%%%%%%%%%%%%

In this section we investigate the convergence properties of the global algorithm.
Figure \ref{fig:convergence_overview_N}  shows the maximal deviation in the gravitational moments
\begin{equation}
    \begin{tabular}{c}
        \text{maximal gravitational} \\ 
        \text{moment deviation}
    \end{tabular}
    = \max_{i \in \{2,4\}} \frac{\left(J^{(k+1)}_{i,\text{ToF}} - J_{i,\text{data}}\right)^2}{\sigma_{i}^2}, 
\end{equation}
versus the maximal relative pressure deviation
\begin{equation}
    \begin{tabular}{c}
        \text{maximal relative} \\ 
        \text{pressure deviation}
    \end{tabular}
    = \max_{r} \frac{\left|P^{(k+1)}_\text{ToF}(r)-P^{(k)}_\text{ToF}(r)\right|}{P^{(k+1)}_\text{ToF}(r)}, 
\end{equation}
for different generations $k\in\{1,3,5\}$.
Each data point represents one model, a complete tuple according to Equation \ref{eq:complete_tuple}.
Points within the hatched area are both consistent with the $\epsilon=0.02$ criterion (Equation \ref{eq:pressure_conistency}) and with the measured gravitational moments within their respective uncertainties (Table \ref{tab:measured_data}).
Such points therefore correspond to physical models.

%%%%%%%%%%%%%%%%%%%%%%%%%%%%%%%%%%%%%%%%%%%%%%%%%%%%%%%%%%%%%%%%%%%%%%%%%%%%%%%%%%%%%%%%%%

Figure \ref{fig:convergence_overview_N} highlights two properties of our algorithm:
First, the histograms show that as we progress to higher generations, our models become more physical.
They align more with the measured gravitational moments and also become more self-consistent according to Equation \ref{eq:pressure_conistency}.
Second, considering the single highlighted path (black arrows) between the generations, the random nature of our compositional algorithm becomes apparent.
While the models become more physical on average, that must not be true for any given model.
The first model (blue point with black circle, $k=1$) has pressure deviations of almost 100\% and gravitational moments more than ten standard deviations away from the measured values.
After just two generations (orange point with black circle, $k=3$), it leads to a physical model thanks to random choices made by the compositional algorithm.
However, two additional generations later (green point with black circle, $k=5$), the result is again far away from the measured gravitational moments.

%%%%%%%%%%%%%%%%%%%%%%%%%%%%%%%%%%%%%%%%%%%%%%%%%%%%%%%%%%%%%%%%%%%%%%%%%%%%%%%%%%%%%%%%%%

We can theretofore conclude that while the non-random parts of the algorithm lead towards physical models on average, the random nature of other algorithm parts ensure that a wide variety of options are explored.
The aforementioned parameter $\epsilon$ should be thought of as a selection criterion, not a convergence criterion.
There is no reason to prefer, for example, a generation $k=5$ models over a generation $k=3$ model.
Higher generations are not necessarily refined versions of what came before.
Finally, Figure \ref{fig:convergence_overview_N} demonstrates that one could easily choose lower values for $\epsilon$.
However, as discussed earlier, lower values for $\epsilon$ would not translate to a closer approach to physical reality.
Each limitation listed in Section \ref{sec:discussion} is capable of introducing deviations comparable to or easily surpassing our $\epsilon=0.02$ threshold. 

%%%%%%%%%%%%%%%%%%%%%%%%%%%%%%%%%%%%%%%%%%%%%%%%%%%%%%%%%%%%%%%%%%%%%%%%%%%%%%%%%%%%%%%%%%
%%%%%%%%%%%%%%%%%%%%%%%%%%%%%%%%%%%%%%%%%%%%%%%%%%%%%%%%%%%%%%%%%%%%%%%%%%%%%%%%%%%%%%%%%%

\section{Comparing different EoS for SiO$_2$}
\label{sec:EoS_comparison}

%%%%%%%%%%%%%%%%%%%%%%%%%%%%%%%%%%%%%%%%%%%%%%%%%%%%%%%%%%%%%%%%%%%%%%%%%%%%%%%%%%%%%%%%%%

Throughout this work, we have emphasised that the EoS employed carry inherent uncertainties, which can lead to discrepancies when compared with alternative EoS formulations.
To illustrate this, Figure \ref{fig:ANEOS1} presents a comparison focused solely on the SiO$_2$ EoS, where we contrast the EoS used in this study \citep{More1988} with the ANEOS formulation for forsterite \citep{Stewart2019}, constructed using M-ANEOS which is based on fitting analytic expressions of the Helmholtz free energy \citep{Thompson1974, Melosh2007, Thompson2019}.

%%%%%%%%%%%%%%%%%%%%%%%%%%%%%%%%%%%%%%%%%%%%%%%%%%%%%%%%%%%%%%%%%%%%%%%%%%%%%%%%%%%%%%%%%%

\renewcommand{\thefigure}{D.\arabic{figure}}

\begin{figure}[!htpb]
    \centering
    \includegraphics[width=\hsize]{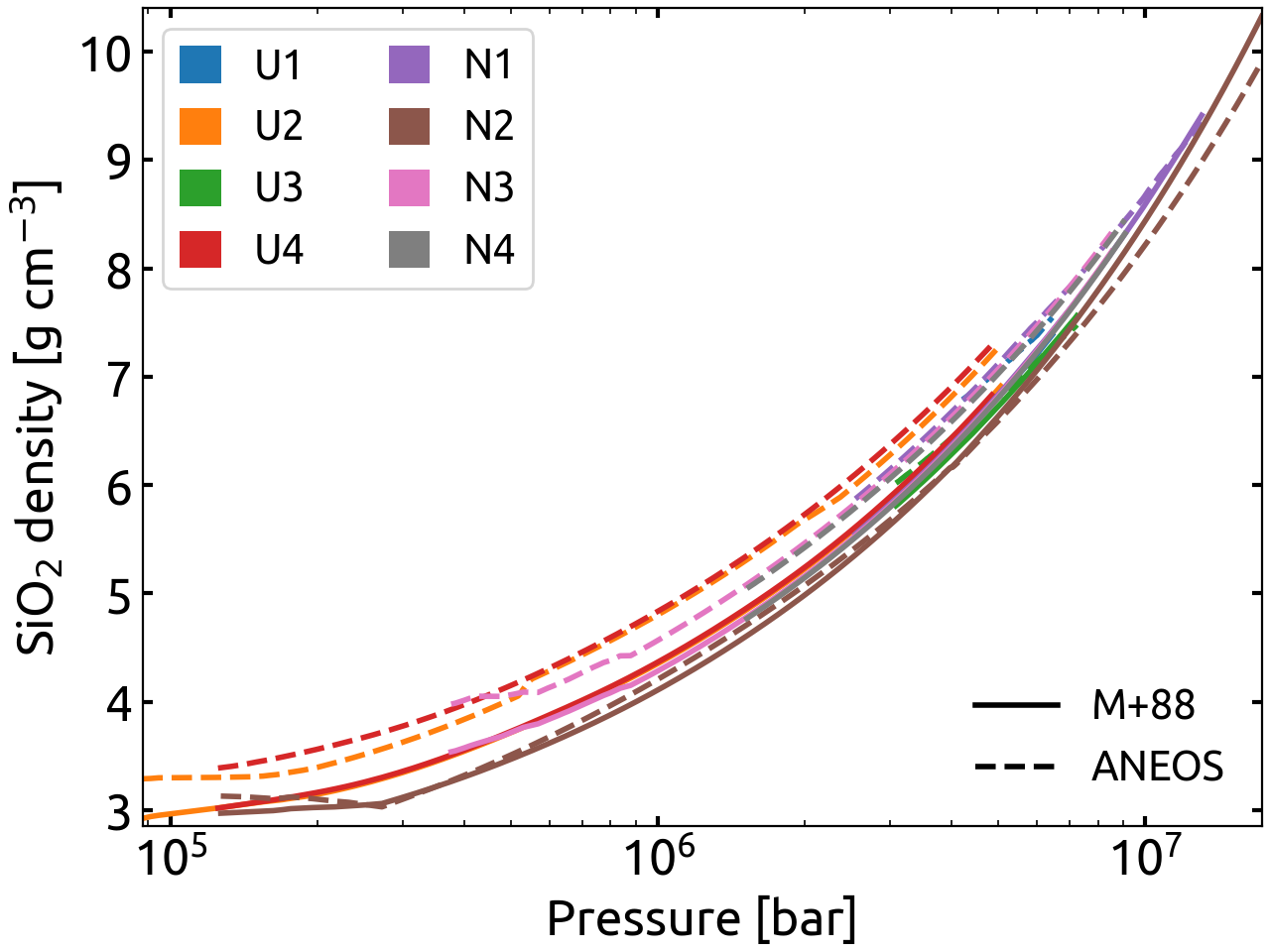}
    \caption{
    Pressure versus density of SiO$_2$ for our Uranus and Neptune models.
    The solid lines correspond to the EoS of \cite{More1988}, while the dashed lines to the ANEOS EoS.
    The density of SiO$_2$ $(\rho_{Z_2})$ is not equal to the model density $(\rho)$ since there are no regions with pure rock in the models (Equation \ref{eq:ideal_mixing}). 
    We only show the regions where SiO$_2$ is present in our models.
    }
\label{fig:ANEOS1}
\end{figure}

%%%%%%%%%%%%%%%%%%%%%%%%%%%%%%%%%%%%%%%%%%%%%%%%%%%%%%%%%%%%%%%%%%%%%%%%%%%%%%%%%%%%%%%%%%

We find that the differences in density between the two SiO$_2$ EoS are generally below 10\%, with many models showing even smaller deviations.
These differences only affect the SiO$_2$ component in our ideal mixing formulation (Equation \ref{eq:ideal_mixing}).
When comparing pressures at constant density instead by inverting the relation, discrepancies in pressure can reach up to 50\%.
Our adopted consistency threshold of $\epsilon = 0.02$ in Equation \ref{eq:pressure_conistency} is clearly conservative in comparison.
Finally, while both EoSs are developed for pure SiO$_2$, the behaviour of materials in mixtures can differ significantly (immiscibility, phase transitions, etc.).
It is therefore desirable to perform detailed simulations of different mixtures at planetary conditions.
Currently, there is no clear justification for preferring one EoS over the other, especially for materials 'heavier' than hydrogen and helium.

%%%%%%%%%%%%%%%%%%%%%%%%%%%%%%%%%%%%%%%%%%%%%%%%%%%%%%%%%%%%%%%%%%%%%%%%%%%%%%%%%%%%%%%%%%
%%%%%%%%%%%%%%%%%%%%%%%%%%%%%%%%%%%%%%%%%%%%%%%%%%%%%%%%%%%%%%%%%%%%%%%%%%%%%%%%%%%%%%%%%%

\end{appendix}

%%%%%%%%%%%%%%%%%%%%%%%%%%%%%%%%%%%%%%%%%%%%%%%%%%%%%%%%%%%%%%%%%%%%%%%%%%%%%%%%%%%%%%%%%%
%%%%%%%%%%%%%%%%%%%%%%%%%%%%%%%%%%%%%%%%%%%%%%%%%%%%%%%%%%%%%%%%%%%%%%%%%%%%%%%%%%%%%%%%%%

\end{document}